\documentclass[ reprint, 
twocolumn,superscriptaddress,
 amsmath,amssymb,
 aps, 
prab
]{revtex4-2}

\usepackage{graphicx}
\usepackage{dcolumn}
\usepackage{bm}
\usepackage{float}
\usepackage{hyperref}
\usepackage[mathlines]{lineno}
\usepackage{multirow}
\usepackage{subcaption} 
\usepackage{color}

\begin{document}

\preprint{APS/123-QED}
\title{\textbf{Alternative Lattice Design for the STCF Collider Rings}}%

\author{Tao Liu}
\affiliation{School of Nuclear Science and Technology, University of Science and Technology of China, No. 443 Huangshan Road, Hefei 230027, P. R. China}

\author{Anton Bogomyagkov}
\affiliation{Budker Institute of Nuclear Physics SB RAS, Novosibirsk 630090, Russia}

\author{Demin Zhou}
\email{dmzhou@post.kek.jp}
\affiliation{KEK, Oho 1-1, Tsukuba, 305-0801, Ibaraki, Japan}
\affiliation{The Graduate University for Advanced Studies (SOKENDAI), Hayama, Kanagawa 240-0193, Japan}

\author{Penghui Yang}
\affiliation{National Synchrotron Radiation Laboratory, University of Science and Technology of China,  No. 42, Hezuohua South Road, Hefei 230029, P. R. China}

\author{Sangya Li}
\affiliation{National Synchrotron Radiation Laboratory, University of Science and Technology of China,  No. 42, Hezuohua South Road, Hefei 230029, P. R. China}

\author{Linhao Zhang}
\affiliation{School of Nuclear Science and Technology, University of Science and Technology of China, No. 443 Huangshan Road, Hefei 230027, P. R. China}

\author{Ye Zou}
\affiliation{School of Nuclear Science and Technology, University of Science and Technology of China, No. 443 Huangshan Road, Hefei 230027, P. R. China}

\author{Jingyu Tang}
\affiliation{School of Nuclear Science and Technology, University of Science and Technology of China, No. 443 Huangshan Road, Hefei 230027, P. R. China}

\author{Qing Luo}
\email{luoqing@ustc.edu.cn}
\affiliation{School of Nuclear Science and Technology, University of Science and Technology of China, No. 443 Huangshan Road, Hefei 230027, P. R. China}
\affiliation{National Synchrotron Radiation Laboratory, University of Science and Technology of China,  No. 42, Hezuohua South Road, Hefei 230029, P. R. China}

\date{\today}

\begin{abstract}

The Super Tau-Charm Facility (STCF) is a proposed high-luminosity electron-positron collider operating in the beam energy range of 1-3.5~GeV, targeting a peak luminosity larger than $0.5\times10^{35}~\mathrm{cm^{-2}s^{-1}}$ at 2~GeV. In this regime, the combination of beam-beam interaction in the crab-waist scheme and low beam energy imposes stringent constraints on dynamic aperture, momentum acceptance, and Touschek lifetime. In this paper, we present an alternative one-fold lattice design for the STCF collider rings, developed within a systematic optimization framework. The approach consists of three stages: (i) lattice-agnostic global parameter optimization using a parameter optimization model that consistently incorporates luminosity performance, beam-beam limits, and collective effects; (ii) optics design based on a compact interaction region with local chromatic correction and crab-waist sextupoles; and (iii) global nonlinear optimization combining analysis-driven methods and tracking-based refinement. The optimized lattice achieves the more ambitious luminosity of $1\times10^{35}~\mathrm{cm^{-2}s^{-1}}$ while maintaining a Touschek lifetime of about 600~s at 2~GeV, with sufficient dynamic aperture and momentum acceptance for stable operation. The results highlight the critical role of local nonlinear control in the interaction region and demonstrate that the proposed optimization strategy provides an effective and general methodology for the design of high-luminosity low-energy colliders.
\end{abstract}

\maketitle

\section{Introduction}

The crab-waist (CW) collision scheme~\cite{Raimondi2006} has revolutionized the design of circular $e^+e^-$ colliders by enabling order-of-magnitude luminosity gains without proportional increases in beam current. Its key principle is to combine a large Piwinski angle with dedicated sextupoles that create a ``crabbed'' waist at the interaction point (IP), thereby suppressing nonlinear beam-beam resonances and enhancing beam stability~\cite{raimondi2007beam, raimondi2008suppression}. First demonstrated successfully at DA$\Phi$NE~\cite{zobov2010TestCrabwaistCollisions}, the CW scheme has since become a cornerstone concept for new-generation factories. It has been exploited at SuperKEKB~\cite{ohnishi2021SuperKEKBOperationUsing} and adopted as a core design strategy in proposed projects such as FCC-ee~\cite{fcc2019fcc}, CEPC~\cite{cepc2018cepc}, and the Super Tau-Charm Facility (STCF)~\cite{peng2020SuperTaucharmFacility, ai2025ConceptualDesignReport}.

Recent design studies in the tau-charm energy region illustrate diverse approaches to implementing the CW collision scheme. Both the Tau/Charm Factory (TCF) in Italy~\cite{biagini2013tau} and the Super Charm-Tau Factory (SCTF) in Russia~\cite{anashin2018super, barnyakov2020project} employ an independent local chromaticity correction system (LCCS) and dedicated crab-waist sextupoles in the interaction region (IR), similar to the former SuperB design~\cite{superb2007superb}. The TCF design featured a short distance of $L^*=0.2~\mathrm{m}$ from the IP to the final doublet (QD0), achieving a large dynamic aperture and long Touschek lifetime. However, the off-axis QD0 layout made the mechanical design and alignment of ancillary systems both complex and costly. Its arc optics adopted a hybrid multi-bend achromat (MBA) structure to achieve low and tunable emittance in a compact ring, and a single Siberian snake (SS) was employed to produce longitudinally polarized electron beam at the IP. In contrast, the SCTF adopted a relatively long $L^*$ of about $0.9~\mathrm{m}$, which is comparable to that of SuperKEKB~\cite{SuperKEKBTDR}. This choice relaxed engineering constraints in the IR, but at the cost of degraded nonlinear performance and a reduced Touschek lifetime. To enable longitudinal electron-beam polarization at the IP, three Siberian snakes were incorporated.

Against this background, the STCF is being developed in China as a high-luminosity $e^+e^-$ collider operating in the tau-charm energy region. 
This facility follows a design philosophy closer to the SCTF approach: it similarly employs independent LCCS and crab-waist sextupoles in the IR. In its earlier design phase, the STCF lattice also adopted three Siberian snakes to enable collisions with a polarized electron beam, mirroring the SCTF scheme. 
However, the lower nominal beam energy of STCF ($2.0~\mathrm{GeV}$, compared with $2.5~\mathrm{GeV}$ for SCTF) further exacerbates challenges related to Touschek lifetime and momentum acceptance, placing more stringent demands on the lattice design.
The design parameter space spans beam energies from 1.0 to 3.5~GeV, with a ring circumference of approximately 800-1000~m. At its nominal operating energy of 2.0~GeV, STCF targets a peak luminosity larger than $0.5\times10^{35} \mathrm{cm}^{-2}\mathrm{s}^{-1}$. To provide sufficient performance margin, the physics design adopts a more ambitious goal of $1.0\times10^{35} \mathrm{cm}^{-2}\mathrm{s}^{-1}$ , which would be unprecedented in this energy range and is the design target pursued in this work. To meet the performance goals, extensive lattice studies have been carried out for STCF~\cite{luo2019progress, lan2021DesignBeamOptics, liu2023DesignHybridSevenbendachromatbased, liu2024RecentProgressFuture, liu2024StudyChromaticityCorrection}, building upon earlier proposals in the same energy domain based on a one-fold ring configuration~\cite{biagini2013tau, anashin2018super}. More recently, the official STCF baseline has evolved toward a nearly two-fold layout to improve optical flexibility and operational robustness, and has moved to an unpolarized beam scheme, 
as documented in~\cite{zou2025optics, ai2025ConceptualDesignReport}.

In this paper, we present a comprehensive design and optimization of the one-fold STCF lattice developed during an earlier stage of the project. Although this configuration predates the current baseline, it captures the essential features of the crab-waist interaction region and associated nonlinear beam dynamics. The main contribution is a two-stage nonlinear optimization strategy, combining analysis-driven pre-optimization with tracking-based refinement, to obtain sufficient dynamic aperture and Touschek lifetime under the constraints of low beam energy and strong crab-waist nonlinearities.

The remainder of this paper is organized as follows. Section~\ref{sec:pom} introduces the design philosophy and the global machine parameters obtained from the parameter optimization model. Section~\ref{sec:ir} presents the IR optics for the STCF rings, together with the underlying theoretical framework for nonlinear control. The design of the technical sections and their integration into the full one-fold ring are described in Sec.~\ref{sec:fullring}. Section~\ref{sec:globaloptimization} details the global nonlinear optimization strategy, including both analysis-driven optimization and tracking-based refinement. Finally, Section~\ref{sec:conclusion} summarizes the main results and outlines directions for future work.

\section{\label{sec:pom} Optimization of machine parameters}

For a high-luminosity tau-charm factory such as STCF, the choice of global machine parameters plays a decisive role in determining whether the target luminosity can be achieved in a sustainable and operationally robust manner. Unlike incremental upgrades of existing facilities, the STCF design must simultaneously satisfy stringent luminosity requirements, unavoidable beam-beam effects inherent to the crab-waist collision scheme, and strong limitations arising from collective effects and beam lifetime at relatively low beam energy. As a result, the optimization of global parameters constitutes a critical first step that precedes detailed lattice design.

For flat-beam collisions with a large Piwinski angle and symmetric beam parameters in a CW collider, the luminosity can be expressed as~\cite{dikansky2009effect, zhou2022formulae}
\begin{equation}
    L=L_0F_h,
\end{equation}
with
\begin{equation}
    L_0=\frac{N_bI_{b+}I_{b-}}{4\pi e^2 f_0 \sigma_x^* \sigma_y^* \sqrt{1+\phi^2}},
    \label{eq:lum}
\end{equation}
\begin{equation}
F_{h} \approx 2\sqrt{\pi}\zeta\exp\left( 4\zeta^{2} \right) \text{Erfc}\left( 2\zeta\right),
\label{eq:Fh}
\end{equation}
where $N_b$ is the number of bunches, $I_{b\pm}$ are the bunch currents, $f_0$ is the revolution frequency, $\sigma_{x,y}^*$ are the transverse beam sizes at the IP, and $\text{Erfc}(x)$ is the complementary error function. The Piwinski angle is defined as $\phi=\sigma_z\tan\theta/\sigma_x^*$ with $\sigma_z$ the bunch length and $\theta$ the half horizontal crossing angle. $F_h$ indicates the hourglass factor with $\zeta=\beta_y^*\tan(2\theta)/(2\sigma_x^*)$. Note that Eq.~\eqref{eq:Fh} is valid with full crab waist applied to the colliding beams.

Crab-waist colliders are typically designed to operate with a large Piwinski angle ($\phi \gg 1$) and a small crossing angle ($\theta \ll 1$), albeit still significantly larger than those used in earlier colliders. In this regime, the geometric luminosity $L_0$ scales with the geometry parameters as $L_0\propto 1/(\sigma_y^*\sigma_z\theta)$. To avoid significant geometric luminosity loss and to suppress strong beam-beam effects associated with the hourglass effects, the condition $\zeta \gtrsim 1$ is typically required, corresponding to $\beta_y^* \gtrsim \sigma_x^*/\theta$. This requirement contrasts with the conventional hourglass condition in earlier colliders, $\beta_y^* \gtrsim \sigma_z$. In the conditions of $\phi\gg 1$ and $\zeta\gtrsim 1$, the beam-beam tune shifts are given by \cite{bbtuneshift}
\begin{equation}
\xi_{y} = \frac{N_{p}r_{e}\beta_{y}^{*}}{2\pi\gamma\sigma_{x}^{*}\sigma_{y}^{*}\sqrt{1 + \phi^{2}}}
\approx \frac{N_{p}r_{e}\beta_{y}^{*}}{2\pi\gamma\sigma_{y}^{*}\sigma_z\theta},
\end{equation}
\begin{equation}
\xi_{x} = \frac{N_{p}r_{e}\beta_{x}^{*}}{2\pi\gamma\sigma_{x}^{*2}\left( 1 + \phi^{2} \right)}
\approx \frac{N_{p}r_{e}\beta_{x}^{*}}{2\pi\gamma\sigma_{z}^{2}\theta^2},
\end{equation}
where symmetric beams are assumed, $N_p$ is the bunch population and $\gamma$ is the Lorentz factor. Guided by empirical beam–beam limits observed in existing circular $e^+e^-$ colliders, the STCF design adopts the constraint $\xi_y \lesssim 0.1$.

In the CW scheme, the luminosity is highly sensitive to perturbations that can increase the vertical beam size at the interaction point, $\sigma_y^*$, which is designed to be extremely small compared with that in earlier collider configurations. The luminosity is controlled by a tightly coupled set involving the beam currents, Twiss functions at the IP, bunch length, crossing angle, and beam-beam tune shifts. These parameters are further constrained by nonlinear beam dynamics, intrabeam scattering (IBS), Touschek scattering, synchrotron radiation damping and excitation, and impedance-driven collective instabilities. An unfavorable choice of any one of these parameters may become a dominant bottleneck, effectively limiting the achievable luminosity even if all others are optimized. Consequently, a global and self-consistent optimization strategy is required, in which luminosity performance and beam-dynamics constraints are treated on an equal footing.

From the luminosity perspective, the key parameters include the vertical beta function at the IP $\beta_y^*$, the vertical emittance $\epsilon_y$, the bunch population $N_p$, the number of bunches, and bunch length $\sigma_{z}$. Achieving a small $\beta_y^*$ is essential for high luminosity, yet it inevitably leads to large beta functions in the final-focus quadrupoles, generating strong chromaticity and enhanced nonlinearities. This, in turn, affects the dynamic aperture and momentum acceptance, directly impacting the Touschek lifetime. At the same time, a small  horizontal beam size at the IP $\sigma_x^*$ is favored to suppress the hourglass effect and increase the Piwinski angle, but it increases the nonlinearities of the IR optics and thus affecting the lifetime.

Longitudinal beam dynamics introduces additional couplings among the momentum compaction factor $\alpha_c$, RF voltage $V_{\mathrm{RF}}$, synchrotron tune $\nu_s$, and energy spread $\sigma_e$. These parameters jointly determine the natural bunch length, RF momentum acceptance, and damping times. In particular, maintaining $\nu_s$ sufficiently larger than the horizontal beam-beam tune shift $\xi_x$ is necessary to avoid luminosity degradation due to horizontal synchro-betatron resonances and coherent X-Z instability~\cite{Ohmi2017PRL}, imposing nontrivial constraints on $\alpha_c$ and RF parameters. Meanwhile, the RF acceptance must exceed the lattice momentum acceptance to avoid RF-induced lifetime limitations.

Single-bunch collective effects further restrict the available parameter space. IBS modifies the transverse emittance, energy spread and bunch length, while Touschek scattering sets a stringent lower bound on the combined transverse beam size, bunch length, and momentum acceptance. Impedance-driven effects such as potential-well distortion, microwave instability (MWI), and transverse mode-coupling instability (TMCI) impose upper limits on the single-bunch current. These effects are particularly critical for STCF due to its relatively low beam energy, where the Touschek lifetime scales unfavorably with energy. Therefore, the global parameter optimization must ensure that the operating point remains safely below instability thresholds while still delivering the target luminosity.

These considerations make the STCF parameter choice a multi-objective problem, in which luminosity, beam-beam limits, collective effects, RF acceptance, dynamic aperture, and Touschek lifetime must be optimized consistently. We therefore employ a lattice-agnostic parameter optimization model (POM) to identify a physically viable region of parameter space before committing to a specific optical implementation. In this model, global parameters such as the ring circumference, emittance, momentum compaction factor, IP beta functions, RF voltage, beam current, and bunch length are varied under constraints derived from luminosity formulas, beam-beam limits, collective-effect thresholds, and damping requirements.

The resulting pre-optimized parameters, summarized in Table~\ref{tab:stcf_params_pom} for the 2-GeV one-fold design, provide the design targets for the lattice developed in the following sections. Although the numerical values may evolve as the baseline design matures, the parameter correlations and trade-offs identified here remain essential guidance for lattice design and nonlinear optimization. For a more detailed discussion of POM, the reader is referred to Ref.~\cite{liu2026nonlinear}.

\begin{table}
\begin{ruledtabular}
\centering
\caption{Machine parameters for the one-fold STCF obtained from lattice-agnostic global parameter optimization.}
\label{tab:stcf_params_pom}
\begin{tabular}{lcl}
\textbf{Parameter} & \textbf{Unit} & \textbf{Value} \\
\hline
Beam energy, $E$ & GeV & 2 \\
Circumference, $C$ & m & 835.12 \\
Full crossing angle, $2\theta$ & mrad & 60 \\
Horizontal emittance, $\epsilon_x$ & nm & 5.9 \\
Coupling factor, $k$ & \% & 0.50 \\
Beta functions at IP, $\beta_x^*/\beta_y^*$ & mm & 40/0.6 \\
Momentum compaction factor, $\alpha_c$ & $\times 10^{-3}$ & 1.3 \\
Energy spread, $\sigma_e$ & $\times 10^{-4}$ & 9.8 \\
Beam current, $I$ & A & 2.02 \\
Particles per bunch, $N_b$ & $\times 10^{10}$ & 5.05 \\
Energy loss per turn, $U_0$ & keV & 400 \\
Damping times, $\tau_x/\tau_y/\tau_z$ & ms & 27.8/27.8/13.9 \\
Harmonic number, $h$ & -- & 1392 \\
RF voltage, $V_{\text{RF}}$ & MV & 1.93 \\
Bunch length, $\sigma_z$ & mm & 11.2 \\
Piwinski angle, $\phi$ & rad & 21.9 \\
Beam-beam parameters, $\xi_x/\xi_y$ & -- & 0.002/0.078 \\
synchrotron tune, $\nu_s$ & -- & 0.016\\
Hourglass factor, $F_h$ & -- & 0.9268 \\
Theoretical peak luminosity, $L$ & cm$^{-2}$s$^{-1}$ & $1.05\times10^{35}$ \\
\end{tabular}
\end{ruledtabular}
\end{table}

\section{\label{sec:ir}Optics of the interaction region for STCF}

Two classes of crab-waist interaction-region optics have been developed. In the local chromatic correction (LCC) scheme~\cite{superb2007superb, raimondi2025local}, the chromaticity generated by the final doublet is corrected locally by dedicated sextupole pairs, leaving the crab-waist sextupoles as independent knobs. In the global hybrid correction (GHC) scheme~\cite{oide2016DesignBeamOptics}, the crab-waist transformation is combined with chromatic correction through a controlled strength imbalance of sextupole pairs. The STCF interaction region adopts the LCC approach because its locality and independent knobs are advantageous for controlling higher-order chromaticity and amplitude-dependent detuning, which are critical for preserving momentum acceptance and dynamic aperture at low beam energy.

\subsection{Modular design for the interaction region}

\begin{figure*}
    \centering
    \includegraphics[width=1\linewidth]{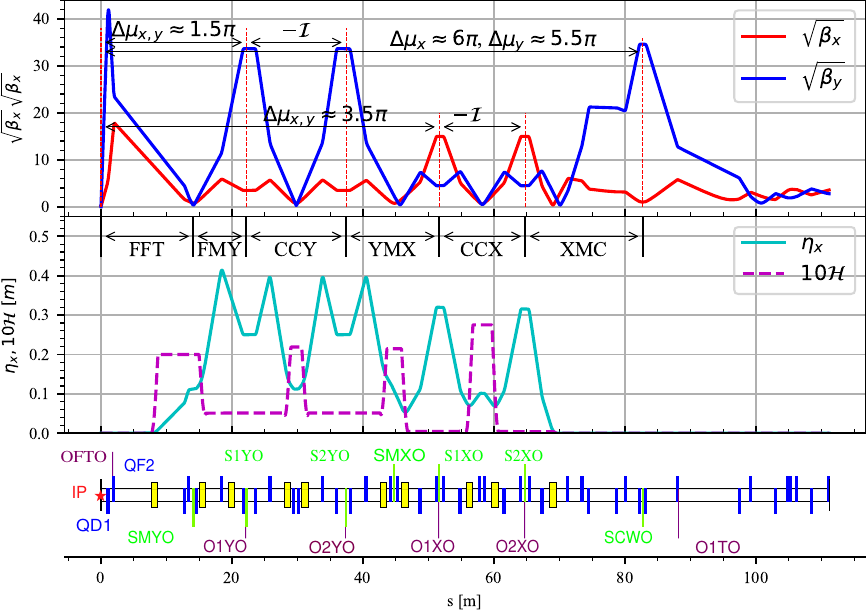}
    \caption{Linear optics and lattice layout of the right-side interaction region for the one-fold STCF. Yellow, blue, and cyan rectangles denote dipoles, quadrupoles, and sextupoles, respectively, while purple lines denote thin octupoles.}
    \label{fig:IRoptics}
\end{figure*}

The one-fold STCF IR employs the LCC scheme, as shown in Fig.~\ref{fig:IRoptics}, and the same modular concept has been adapted to the two-fold configuration~\cite{zhang2025CrabwaistInteractionRegion}. 
From the IP to the right-side crab sextupole (SCWO), the optics is arranged as: final-focus telescope (FFT) $\rightarrow$ matching to vertical chromatic correction (FMY) $\rightarrow$ vertical chromatic correction (CCY) $\rightarrow$ matching (YMX) $\rightarrow$ horizontal chromatic correction (CCX) $\rightarrow$ matching to the crab sextupoles (XMC). On the left side of the IP, the optics follows a similar structure, with slight modifications in the bending magnets to accommodate the finite crossing angle geometry. In the remainder of this section, we first discuss the physical principles underlying the nonlinear optimization, and then describe the multipole configurations used for local nonlinear correction.

A central limitation of CW colliders is the interplay between the crab sextupole pair and the nonlinearities accumulated between them~\cite{bogomyagkov2016FinalFocusDesigns, raimondi2025local}. Although the leading geometric terms cancel for particles close to the design orbit under appropriate phase conditions~\cite{brown1985FirstSecondorderCharged}, this cancellation is degraded by momentum deviation and large betatron amplitudes. Robust performance therefore requires the optics between the crab sextupoles to remain nearly transparent, i.e., weakly dependent on both momentum and amplitude. This motivates the two complementary optimizations discussed below: momentum-dependent chromatic control and amplitude-dependent detuning control.

\subsection{Optimization of momentum-dependent nonlinearities}

Momentum-dependent nonlinearities are characterized by the tune expansion
\begin{equation}
\nu_u=\nu_{u0}+\xi_{u1}\delta+\xi_{u2}\delta^2+\xi_{u3}\delta^3,
\end{equation}
where $u\in\{x,y\}$, $\nu_{u0}$ is the nominal tune, and $\xi_{u1,2,3}$ denote the first- to third-order chromaticities. 
Including contributions from quadrupoles, sextupoles, and octupoles, the chromaticities through third order can be written as~\cite{bengtsson1997sextupole, bogomyagkov2016chromaticity}
\begin{equation}
    \xi_{u1}=  \frac{\delta_u}{4 \pi} \int_0^C \beta_u(s)G_1(s) d s,\label{eq:firstchrom}
\end{equation}
\begin{align}
    \xi_{u2}=  &- \xi_{u1}-\frac{\delta_u}{4 \pi} \int_0^C \beta_u(s)G_2(s) ds \nonumber \\
    &+ \frac{\delta_u}{8 \pi} \int_0^C \beta_u(s) B_{u1}(s)G_1(s) d s,\label{eq:secondchrom}
\end{align}
\begin{align}
    \xi_{u3} =  & \xi_{u1}-\frac{\delta_u}{4 \pi} \int_0^C \beta_u(s)W_{u1}^2(s)G_1(s) d s \nonumber \\ 
    & +\frac{\delta_u}{4 \pi} \int_0^C \beta_u(s)(G_2(s)+G_3(s)) d s \nonumber \\
    & +\frac{\delta_u}{4 \pi} \int_0^C \beta_u(s) B_{u2}(s)G_1(s) d s,\label{eq:thirdchrom}
\end{align}
with $\delta_u=-1$ for $u=x$ and $\delta_u=1$ for $u=y$. The gradient-dependent functions are
\begin{equation}
    G_1(s)=K_1(s)-K_2(s) \eta_{x0}(s),
\end{equation}
\begin{equation}
    G_2(s)= K_2(s) \eta_{x1}(s)
    + \frac{1}{2} K_3(s) \eta_{x0}^2(s),
\end{equation}
\begin{equation}
    G_3(s)= -K_2(s)\eta_{x2}(s)-K_3(s)\eta_{x0}(s)\eta_{x1}(s),
\end{equation}
where $K_{1,2,3}(s)$ are the normalized strengths of quadrupoles, sextupoles, and octupoles, respectively. 
The horizontal orbit is expanded as
\begin{equation}
    x(s,\delta)=\eta_{x0}(s)\delta+\eta_{x1}(s)\delta^2+\eta_{x2}(s)\delta^3,
\end{equation}
defining the dispersion functions up to second order.

Off-momentum optical distortions enter Eqs.~\eqref{eq:secondchrom}--\eqref{eq:thirdchrom} through the Montague functions~\cite{montague1979linear}. The first-order Montague $W$ function is
\begin{equation}
    W_{u1}(s)=\sqrt{A_{u1}^2(s) + B_{u1}^2(s) },
    \label{eq:montague}
\end{equation}
with
\begin{equation}
    A_{u1}=\left. \frac{\partial\alpha_{u}}{\partial\delta}-\frac{\alpha_{u}}{\beta_{u}}\frac{\partial\beta_{u}}{\partial\delta} \right |_{\delta=0}, \qquad
    B_{u1}=\left. \frac{1}{\beta_{u}}\frac{\partial\beta_{u}}{\partial\delta}\right |_{\delta=0},
    \label{eq:ab1}
\end{equation}
and explicit forms
\begin{equation}
    A_{u1}(s)= -\frac{\delta_u}{2\sin(2\pi\nu_u)} \int_0^{C} \beta_{u}(s') G_1(s') 
\sin \bigl(\phi_{ss'}\bigr)\,d s', \label{eq:alphabeating}
\end{equation}
\begin{equation}
    B_{u1}(s)=-\frac{\delta_u}{2\sin(2\pi\nu_u)} \int_0^{C} \beta_{u}(s') G_1(s') 
\cos \bigl(\phi_{ss'}\bigr)\,d s', \label{eq:betabeating}
\end{equation}
where $\phi_{ss'}=2\pi\nu_u-2\left|\mu_{u}(s)-\mu_{u}(s')\right|$ and $\mu_u$ is the phase advance. 
Second-order analogs $A_{u2},B_{u2}$ are defined as~\cite{cai2022OptimizationChromaticOptics, bogomyagkov2024thesis}
\begin{equation}
A_{u2}=\left. \frac{1}{2}\left[\frac{\partial^2 \alpha_{u}}{\partial \delta^2}-\frac{\alpha_{u}}{\beta_{u}} \frac{\partial^2 \beta_{u}}{\partial \delta^2}\right] \right |_{\delta=0}, \
B_{u2}=\left. \frac{1}{2\beta_{u}} \frac{\partial^2 \beta_{u}}{\partial \delta^2} \right |_{\delta=0}.
\label{eq:ab2}
\end{equation}
We follow~\cite{cai2022OptimizationChromaticOptics} to take Eqs.~\eqref{eq:ab1} and~\eqref{eq:ab2} as Montague functions.
In the present work, the off-momentum beta-beating terms are retained up to second order, which is sufficient to describe the dominant distortions relevant for the present IR optimization; higher-order extensions exist but are not required here~\cite{cai2022OptimizationChromaticOptics, bogomyagkov2024thesis}.
For simplicity, the bending-magnet contribution to chromaticity is neglected in the analytical discussion below.

\begin{figure}[h]
    \centering
    \includegraphics[width=1\linewidth]{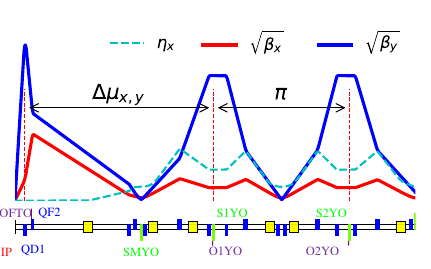}
    \caption{Schematic layout of the FD quadrupoles (QD1/QF2) and the CCY sextupole pair (S1YO/S2YO) in the one-fold STCF IR. Here, $\Delta\mu_{x,y}$ denotes the horizontal and vertical phase advances from QD1 to S1YO.}
    \label{fig:FD_CCY_for_chrom}
\end{figure}

\paragraph*{Local chromaticity compensation in the IR.}
The dominant IR chromaticity source is the final doublet of the quadrupoles (FD). 
To illustrate the local correction principle, we consider the FFT-CCY segment in Fig.~\ref{fig:FD_CCY_for_chrom}. 
On the right of the IP, the FD is formed by QD1 and QF2; the vertical chromaticity is corrected locally by the CCY sextupole pair S1YO/S2YO, arranged at a phase advance of $\pi$ to cancel the leading geometric terms~\cite{brown1985FirstSecondorderCharged}. 
Neglecting other elements, the first-order contributions from FD quadrupoles and CCY sextupoles follow from Eq.~\eqref{eq:firstchrom}:
\begin{equation}
\xi_{yQ}=\sum_i \bigl\langle \beta_y K_1 \bigr\rangle_{Q_i},\quad
\xi_{yS}=\sum_j \bigl\langle \beta_y K_2 \eta_{x0} \bigr\rangle_{S_j},
\end{equation}
where $\langle\cdot\rangle$ denotes integration over finite magnet length. 
Local cancellation requires $\xi_{yQS}\equiv \xi_{yQ}-\xi_{yS}=0$. 
In CW optics, the very large $\beta$ functions in the FD imply strong sextupoles for local compensation, which makes controlling higher-order chromaticity and detuning essential.

\paragraph*{Phase constraints for higher-order chromaticity.}
Second- and third-order chromaticities depend on the dispersions and off-momentum beta beating through Montague functions in Eqs.~\eqref{eq:secondchrom}--\eqref{eq:thirdchrom}. Because the FD produces both large $\beta$ and a relatively small phase advance between QD1 and QF2, it is beneficial to choose the phase advance from QD1 to the CCY sextupoles close to an integer multiple of $\pi$ (see Fig.~\ref{fig:FD_CCY_for_chrom}), yielding the approximation
\begin{equation}
B_{y1}(s)\approx
-\frac{\cos\Pi_y(s)}{2\sin(2\pi\nu_y)} \,\xi_{yQS},
\end{equation}
with $\Pi_y(s)=2\bigl(\pi\nu_y-|\mu_{yQ1}-\mu_y(s)|\bigr)$ and $\mu_{yQ1}$ the phase at QD1. 
Under the conditions $\xi_{yQS}=0$ and $\Delta\mu_y\simeq k\pi$, the residual second-order chromaticity is dominated by second-order dispersion:
\begin{equation}
\xi_{y2}\simeq -\xi_{y1}-\frac{1}{4\pi}\sum_j \bigl\langle \beta_y K_2 \eta_{x1} \bigr\rangle_{S_j},
\end{equation}
and, analogously, $\xi_{y3}$ is governed mainly by $\eta_{x1}$ and $\eta_{x2}$ when the local first-order cancellation and the phase constraint are satisfied.

\paragraph*{Sensitivity to phase-advance deviations.}
Deviations from the ideal phase relations (see Fig.~\ref{fig:FD_CCY_for_chrom}), between QD1 and QF2 ($\Delta\mu_{yQQ}$), between QD1 and S1YO ($\Delta\mu_{yQS}$), and between S1YO and S2YO ($\Delta\mu_{ySS}$), perturb the chromaticities. 
In design, these deviations can be used as weak knobs to redistribute higher-order chromaticity; in operation, they must be tightly controlled and/or implemented as beam-based tunable knobs. 
For small $\Delta\mu_{yQS}$, the induced variations are approximately
\begin{equation}
    \Delta \xi_{y2} \approx
    \frac{\Delta\mu_{yQS}}{4\pi} \xi_{yQ} \xi_{yS},
\end{equation}
\begin{align}
    \Delta \xi_{y3} \approx
     &-\frac{\Delta\mu_{yQS}}{2\pi}  \xi_{yQ} \xi_{yS}
    + \frac{\Delta\mu_{yQS}}{4\pi} \xi_{yQ} \sum_j \bigl< \beta_yK_2\eta_{x1} \bigr>_{Sj} \nonumber \\
    &-\frac{\Delta\mu_{yQS}}{2\pi\tan(2\pi\nu_y)} \xi_{yQ}\xi_{yS}\xi_{yQS}.
\end{align}
Similarly, small deviations inside the FD and inside the CCY pair produce second-order contributions
\begin{equation}
    \Delta\xi_{y2}=
    \frac{\Delta\mu_{yQQ}}{4\pi} \left( \xi_{yQS}-\bigl\langle\beta_yK_1\bigr\rangle_{Q2} \right) \bigl\langle\beta_yK_1\bigr\rangle_{Q2},
\end{equation}
\begin{equation}
    \Delta\xi_{y2}=
    \frac{\Delta\mu_{ySS}}{4\pi} \left( \xi_{yQS}-\bigl\langle\beta_yK_2\eta_{x0}\bigr\rangle_{S2} \right) \bigl\langle\beta_yK_2\eta_{x0}\bigr\rangle_{S2}.
\end{equation}
The above expressions are presented for the vertical plane; analogous formulations can be derived for the horizontal plane and are omitted here for brevity. These relations highlight a general feature of CW IRs: the very large $\beta$ functions at the FD and at the local chromatic sextupoles amplify the chromatic sensitivity to phase errors, making phase control a critical design and operational requirement.

\paragraph*{Practical considerations.}
The discussion above isolates the FD-CCY subsystem to expose the basic correction mechanism. In the full IR, additional quadrupoles and matching sections modify these idealized conditions and contribute to chromaticity. These effects are included in the full IR optimization and global matching of the STCF ring optics. Additional mirror-symmetric sextupoles, such as SMYO and SMXO in Fig.~\ref{fig:IRoptics}, are introduced to provide further flexibility for higher-order chromatic correction~\cite{raimondi2025local}.

\subsection{Optimization of amplitude-dependent nonlinearities}\label{sec:ampl_dependent}

Amplitude-dependent nonlinearities primarily appear as nonlinear tune shifts, which degrade the effective cancellation of the crab sextupole pair and reduce the dynamic aperture. 
In the IR, the dominant contribution is typically the first-order amplitude-dependent tune shift (ADT), so the optimization targets the leading detuning coefficients:
\begin{equation}
\label{eq:amplitude_tuneshift}
\begin{aligned}
& \Delta \nu_x=\alpha_{xx} J_x+\alpha_{xy} J_y, \\
& \Delta \nu_y=\alpha_{yx} J_x+\alpha_{yy} J_y,
\end{aligned}
\end{equation}
with $\alpha_{xy}\simeq\alpha_{yx}$ in most practical cases. 
In CW IRs, the main contributions arise from: (i) the kinematic term in the final drift, (ii) Maxwellian fringe fields of the FD quadrupoles, and (iii) finite-length effects of strong $-\mathcal{I}$ sextupole pairs~\cite{bogomyagkov2024TouschekLifetimeLuminosity}. 
Below we employ the analytical expressions of Ref.~\cite{bogomyagkov2024TouschekLifetimeLuminosity} and apply them directly to the STCF optics.

\paragraph*{Kinematic term in the final drift.}
For the drift between the IP and QD1 (length $L^*$), modeling QD1 as a thin lens at its center adds $L_q/2$ to the effective drift length and yields \cite{bogomyagkov2016FinalFocusDesigns,bogomyagkov2024thesis}
\begin{equation}
\label{eq:ADTs_kinematic}
\begin{aligned}
\alpha_{x x} & =\frac{3}{16 \pi} \oint \gamma_x^2(s) d s \approx \frac{3}{16 \pi} \frac{\left(L^{*}+\frac{L_q}{2}\right)}{\beta_{x}^{*2}}, \\
\alpha_{x y}& =\frac{1}{8 \pi} \oint \gamma_x(s) \gamma_y(s) d s \approx \frac{1}{8 \pi} \frac{\left(L^{*}+\frac{L_q}{2}\right)}{\beta_{x}^{*} \beta_{y}^{*}}, \\
\alpha_{y y} & =\frac{3}{16 \pi} \oint \gamma_y^2(s) d s \approx \frac{3}{16 \pi} \frac{\left(L^{*}+\frac{L_q}{2}\right)}{\beta_{y}^{*2}},
\end{aligned}
\end{equation}
where $\gamma_{x,y}(s)$ denote the Twiss parameters along the beam line and $\beta_{x,y}^{*}$ are the beta functions at the IP. Because CW colliders require very small $\beta_{x,y}^*$, this contribution can be significant even when $L^*$ is constrained by the machine-detector interface~\cite{zhang2025CrabwaistInteractionRegion}.

\paragraph*{FD fringe fields.}
Fringe-field contributions from the FD can be written as~\cite{bogomyagkov2024TouschekLifetimeLuminosity}
\begin{equation}
\label{eq:ADTs_fringe}
\begin{aligned}
\alpha_{x x} & = -\frac{1}{32 \pi} \oint K_1^{\prime 2} \beta_x^2 d s \approx -\frac{1}{8 \pi}\sum_i\delta_{i}K_{1, i}\alpha_{x, i} \beta_{x, i}, \\
\alpha_{x y} & = \frac{1}{8 \pi} \oint K_1^{\prime}\left(\alpha_y \beta_x -\alpha_x \beta_y\right) d s \\
& \approx \sum_i \frac{\delta_{i}K_{1, i}}{8 \pi}\left(\alpha_{y, i} \beta_{x, i}-\alpha_{x, i} \beta_{y, i}\right), \\
\alpha_{y y} & = \frac{1}{32 \pi} \oint K_1^{\prime \prime} \beta_y^2 d s \approx \frac{1}{8 \pi}\sum_i \delta_{i} K_{1, i}\alpha_{y, i} \beta_{ y, i},\\
\end{aligned}
\end{equation}
where $K_{1,i}$ denotes the normalized gradient of the quadrupole adjacent to the $i$-th fringe field, $\alpha_{(x,y),i}$ and $\beta_{(x,y),i}$ are the corresponding Twiss parameters, and $\delta_i=1$ ($-1$) denotes the entrance (exit) fringe, respectively. 
The exceptionally large $\beta$ functions and strongly asymmetric edge optics of the FD make these terms particularly important, often dominating the cross-plane detuning.

\begin{figure}[h]
    \centering
    \includegraphics[width=1\linewidth]{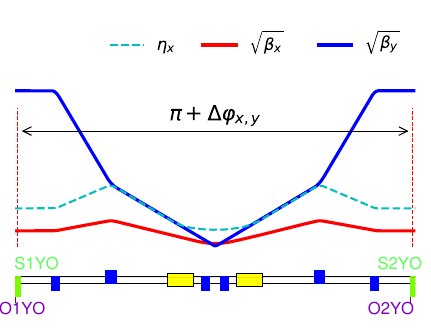}
    \caption{Schematic of the thick sextupole pair S1YO/S2YO connected by a non-exact $-\mathcal{I}$ transformation. The phase advances from the center of S1YO to the center of S2YO are adjusted to $\pi+\Delta\varphi_{x,y}$ in the horizontal and vertical planes.}
    \label{fig:STCF_minus_I}
\end{figure}

\paragraph*{$-\mathcal{I}$ sextupole pairs and detuning knobs.}
A thick $-\mathcal{I}$ sextupole pair with small phase errors $\Delta\varphi_{x,y}$, as illustrated in Fig.~\ref{fig:STCF_minus_I},  contributes to detuning as~\cite{bengtsson1997sextupole, wei2023MinimizingFluctuationResonance}
\begin{equation}
\label{eq:ADTs_sext_pair}
\begin{aligned}
\alpha_{xx} & = -\frac{(K_2 L)^2\beta_{\mathrm{x}}^2 \left( L-3\Delta\varphi_x\beta_{\mathrm{x}} \right)}{16 \pi } ,\\
\alpha_{xy} & = -\frac{(K_2 L)^2 \beta _{\mathrm{x}} \beta _{\mathrm{y}}\left( L-3(\Delta\varphi_x\beta_{\mathrm{x}}-2\Delta\varphi_y\beta_{\mathrm{y}} ) \right)}{24 \pi } ,\\
\alpha_{yy} & = -\frac{(K_2 L)^2\beta_{\mathrm{y}}^2 \left( L-3\Delta\varphi_x\beta_{\mathrm{x}} \right)}{16 \pi } ,
\end{aligned}
\end{equation}
where $K_2$, $L$ are the sextupole strength and length. 
This relation clarifies how small phase adjustments in CCY/CCX can selectively tune ADT components (notably $\alpha_{xy}$ and $\alpha_{yy}$ in regions with $\beta_y\gg\beta_x$), as suggested in Ref.~\cite{raimondi2025local}.

\paragraph*{Octupoles.}
A flexible and largely orthogonal control of detuning is provided by dedicated octupoles:
\begin{equation}
\label{eq:ADTs_octu}
\begin{aligned}
\alpha_{x x} & =\frac{1}{16 \pi} \oint d s\, K_3(s) \beta_x^2(s), \\
\alpha_{x y} & =-\frac{1}{8 \pi} \oint d s\, K_3(s) \beta_x(s) \beta_y(s), \\
\alpha_{y y} & =\frac{1}{16 \pi} \oint d s\, K_3(s) \beta_y^2(s),
\end{aligned}
\end{equation}
where $K_3(s)$ is the octupole strength. 
The tuning efficiency is therefore dictated by the local optics: regions with $\beta_y\gg\beta_x$ are optimal for $\alpha_{yy}$ control, $\beta_x\gg\beta_y$ for $\alpha_{xx}$, and comparable $\beta_x,\beta_y$ for $\alpha_{xy}$. 
For an LCC IR, this naturally suggests octupole locations near CCY (see Fig.~\ref{fig:STCF_minus_I}, to tune $\alpha_{yy}$), near CCX (to tune $\alpha_{xx}$), and near the FD (to tune $\alpha_{xy}$).

\subsection{Application to the STCF IR}

Following the principles discussed above, the STCF IR implements a compact LCC-based CW layout with dedicated local knobs for chromaticity correction, crab-waist transformation, and detuning control. Referring to Fig.~\ref{fig:IRoptics} for the right-side IR, the sextupole pairs S1YO/S2YO and S1XO/S2XO provide vertical and horizontal chromaticity correction, respectively. The mirror-symmetric sextupoles SMYO and SMXO are used for third-order chromaticity correction (see Eq.~\eqref{eq:thirdchrom}). The octupole pairs O1YO/O2YO and O1XO/O2XO, placed near the chromaticity-correction sextupoles, provide local control of the dominant ADTs in the vertical and horizontal planes. The octupole OFTO near QF2 is used to correct the coupling ADTs $\alpha_{xy}$ and $\alpha_{yx}$ within the crab-sextupole pair, while O1TO outside the crab-sextupole pair provides an additional knob for global nonlinear optimization.

Chromatic optimization follows the analytical considerations outlined previously. The phase advances between FD and CCY/CCX are set close to integer multiples of $\pi$ to suppress chromatic beta beating at the correction sextupoles, while each CCY/CCX pair is arranged close to form a $-\mathcal{I}$ transformation to cancel leading geometric aberrations. 
Within these constraints, the dispersions at the LCC sextupoles are chosen by balancing sextupole strength against off-momentum acceptance. 
Larger dispersion reduces the required sextupole strength and hence geometric aberrations (see Eq.~\eqref{eq:ADTs_sext_pair}), and also relaxes sensitivity to orbit errors~\cite{raimondi2025local}; however, it increases the dispersion invariant
\begin{equation}
\mathcal{H}=\gamma_x\eta_x^2+2\alpha_x\eta_x\eta_x'+\beta_x\eta_x'^2,
\end{equation}
which enhances off-momentum distortions and can limit the local momentum acceptance (LMA) and Touschek lifetime. 
The STCF solution therefore adopts a moderate dispersion level that preserves both practical sextupole strengths and sufficient momentum acceptance.

The CW phase conditions between the IP and the crab sextupoles are imposed in the standard form,
\begin{equation}
\Delta\mu_x = m\pi,\qquad \Delta\mu_y = \left(n+\frac{1}{2}\right)\pi,
\end{equation}
with integers $m,n$, while the surrounding matching sections are used to minimize the momentum and amplitude dependence of the optical functions at the crab sextupoles. 
The resulting octupole configuration follows the optics-based placement criteria discussed in Sec.~\ref{sec:ampl_dependent}. In the present design, $-\mathcal{I}$ phase tuning is not used in routine optimization in order to keep the on-momentum optics simple; the primary detuning control is instead provided by the dedicated octupoles.

\section{\label{sec:fullring}Optics of technical sections for the one-fold STCF}

\subsection{Layout}
Following the IR design described in the previous section, the remaining sections of the one-fold STCF ring are designed to connect the IR, provide the required radiation damping and polarization control, and preserve the optical conditions needed for global nonlinear optimization.

\begin{figure}[h]
    \centering
    \includegraphics[width=1\linewidth]{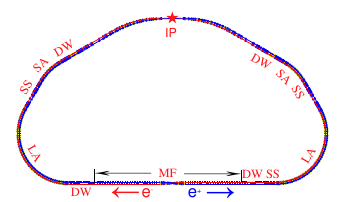}
    \caption{Layout of the one-fold STCF collider rings.}
    \label{fig:layout}
\end{figure}

\begin{figure*}[ht!]
\centering
\includegraphics[width=0.9\linewidth]{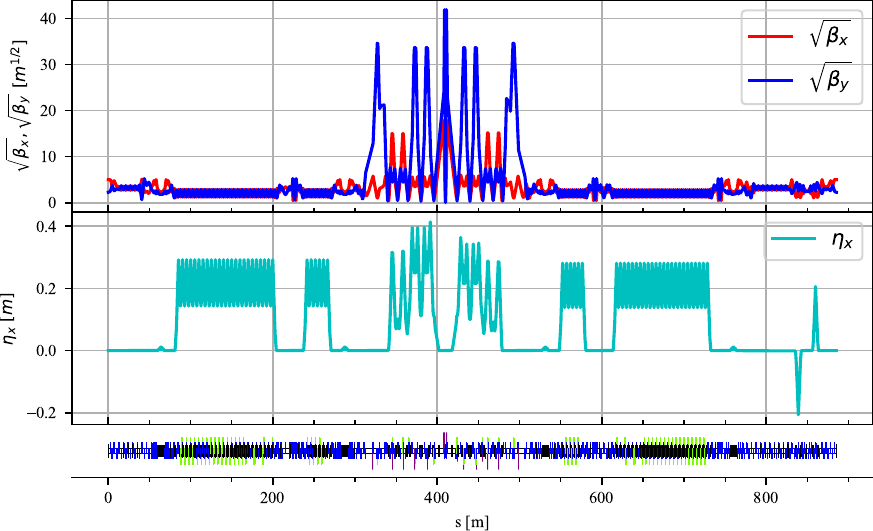}
\caption{Full-ring linear optics and lattice layout of the one-fold STCF, with $s=0$ corresponding to the injection point.}
    \label{fig:ring_optics}
\end{figure*}

The lattice design is based on the pre-optimized global parameters listed in Table~\ref{tab:stcf_params_pom}. As shown in Fig.~\ref{fig:layout}, the collision ring consists of one IR, four arc sections including two long arc (LA) sections and two short arc (SA) sections, three Siberian Snake (SS) sections, four damping wiggler (DW) sections, and one multi-function (MF) region. The positron ring is geometrically symmetric to the electron ring. The dual-ring configuration is realized by inner and outer half-rings that differ only in dipole curvature radii within the arc sections, while all other complex units share identical optical structures.
The full linear optics functions of the collision ring are presented in Fig.~\ref{fig:ring_optics}. The detailed design of these sections is discussed in the following 
subsections. After integrating and jointly optimizing all sections under 
these constraints, the resulting machine parameters for the one-fold STCF are summarized in 
Table~\ref{tab:stcf_lat_params}.
\begin{table}
\begin{ruledtabular}
\centering
\caption{Optimized parameters for the one-fold alternative design of the STCF}
\label{tab:stcf_lat_params}
\begin{tabular}{lcl}
\textbf{Parameter} & \textbf{Unit} & \textbf{Value} \\
\hline
Beam energy, $E$ & GeV & 2 \\
Circumference, $C$ & m & 885.232 \\
Beam current, $I$ & A & 2 \\
Full crossing angle, $2\theta$ & mrad & 60 \\
Beta functions at IP, $\beta_x^*/\beta_y^*$ & mm & 40/0.6 \\
Coupling factor, $k$ & \% & 0.50 \\
Betatron tunes $\nu_{x}/\nu_{y}$ & & 33.538/33.58 \\
Horizontal emittance, $\epsilon_x$ (SR/IBS) & nm & 5.77/6.22 \\
Synchrotron tune, $\nu_{s}$ & & 0.0165 \\
Momentum compaction factor, $\alpha_c$ & $\times 10^{-3}$ & 1.18 \\
Energy spread, $\sigma_e$ (SR/IBS) & $\times 10^{-4}$ & 9.70/9.91 \\
Particles per bunch, $N_b$ & $\times 10^{10}$ & 5.0 \\
Energy loss per turn, $U_0$ & keV & 382.7 \\
Damping times, $\tau_x/\tau_y/\tau_z$ & ms & 30.6/30.8/15.4 \\
Harmonic number, $h$ & -- & 1476 \\
RF voltage, $V_{\text{RF}}$ & MV & 2.0 \\
Bunch length, $\sigma_z$ (SR/IBS) & mm & 9.76/9.99 \\
RF energy acceptance, $\delta _{\text{RF}}$ & \% & 1.62 \\
Piwinski angle, $\phi$ &  & 19.0 \\
Beam-beam parameters, $\xi_x/\xi_y$ & -- & 0.0025/0.084 \\
Hourglass factor, $F_h$ & -- & 0.9235 \\
Theoretical peak luminosity, $L$ & cm$^{-2}$s$^{-1}$ & $1.12\times10^{35}$ \\
\end{tabular}
\end{ruledtabular}
\end{table}

\subsection{Arc Sections}

The arc sections are designed to realize the dual-ring geometry within a common tunnel. To accommodate the dual-ring configuration, each ring is divided into inner and outer half-rings. Each half-ring consists of one long arc with a total bending angle of $120^\circ$ and one short arc with a bending angle of $30^\circ$, as illustrated in Fig.~\ref{fig:layout}. The choice of this deflection angle ensures uniform azimuthal distribution of the three Siberian Snakes, which is necessary to achieve longitudinal polarization of the electron beam at the IP.

\begin{figure}[h]
\centering
\includegraphics[width=1\linewidth]{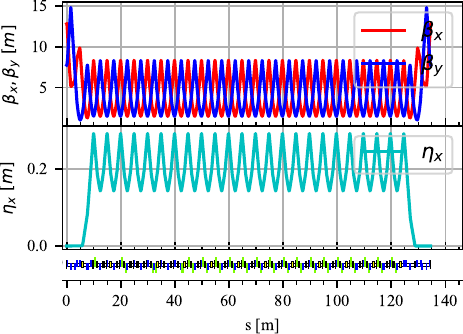}
\caption{Linear optics and lattice layout of the long-arc section in the outer half-ring of the one-fold STCF.}
    \label{fig:outerlongarc}
\end{figure}

\begin{figure}[h]
\centering
\includegraphics[width=1\linewidth]{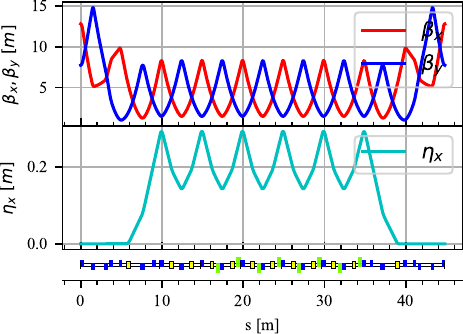}
\caption{Linear optics and lattice layout of the short-arc section in the outer half-ring of the one-fold STCF.}
    \label{fig:outershortarc}
\end{figure}

Both long and short arcs are constructed from FODO cells with horizontal 
and vertical phase advances of $90^\circ$ per cell and a dipole bending angle of $5^\circ$. The inner and outer arcs share the same layout, differing only in the dipole curvature radii. The inner half-ring uses dipoles with a curvature radius of 10~m, whereas the outer half-ring employs a curvature radius of 12.14~m. Their centers coincide, so that the transverse separation between the electron and positron beam centers remains constant at 2.14~m throughout the arc sections. Quadrupole, sextupole, and drift lengths are kept identical in both half-rings in order to preserve symmetry and simplify correction strategies. Figures~\ref{fig:outerlongarc} and~\ref{fig:outershortarc} present the lattice and corresponding linear optics functions of the long and short arc sections in the outer half-ring.

The short arc region consists of five FODO cells. Among them, four consecutive FODO cells are configured such that a sextupole is placed downstream of each quadrupole for correcting both horizontal and vertical chromaticity, and these four FODO cells are completely identical (including the strengths of their sextupoles). This configuration represents a balance between mitigating nonlinear resonances and correcting chromaticity. A high-filling scheme is beneficial for locally correcting the chromaticity of the IR matching section and DW sections as much as possible, which helps achieve good off-momentum dynamics performance. These four identical FODO cells form a Higher-Order Achromat (HOA) structure~\cite{brown1985FirstSecondorderCharged}, in which 
first- and second-order geometric driving terms largely cancel. This arrangement significantly reduces nonlinear resonance excitation while maintaining strong chromatic correction capability. The nonlinearity configured in this way is better than that in the case where sextupoles are installed downstream of quadrupoles in all five FODO cells.

The long arc contains 23 FODO cells and is organized into two functional 
regions. The first seven cells employ three non-interleaved 
$-\mathcal{I}$ sextupole pairs for chromaticity correction and Montague
$W$-function control. Since the dispersion function is smaller near 
locations of maximum vertical $\beta$, two sextupole pairs are dedicated 
to vertical chromaticity correction and one pair to horizontal correction, 
thereby improving correction efficiency while limiting nonlinear side effects. In the remaining sixteen cells, a higher sextupole filling scheme is adopted, with one sextupole placed downstream of each quadrupole. These sextupoles form interleaved $-\mathcal{I}$ pairs. Every four identical FODO cells also constitute a HOA structure. In this region, the Montague $W$ function oscillates periodically and is approximately 
matched at the entrance and exit of the module, preserving optical 
transparency to adjacent sections.

Both long and short arcs are connected to the straight sections through 
achromatic matching sections.

\subsection{Multi-function Section}

The multi-function region is the section located opposite 
the IR (see Fig.~\ref{fig:layout}), with a total deflection angle of zero. It serves as the primary operational section of the ring, providing both geometric connection and optical matching between the inner and outer half-rings, while accommodating essential subsystems such as RF cavities and injection elements.

\begin{figure}[h]
\centering
    \includegraphics[width=1\linewidth]{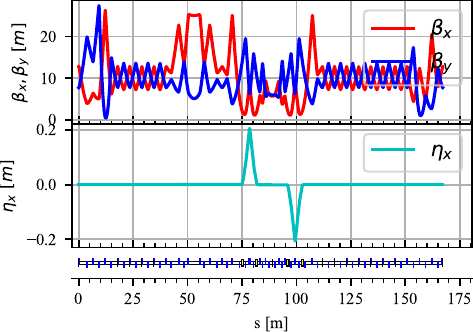}
\caption{Linear optics and lattice layout of the multi-function section for the one-fold STCF.}
    \label{fig:techoptics}
\end{figure}

From the optics perspective, this region is configured to offer flexibility in tuning the working point over a wide range, with minimal impact on amplitude-dependent detuning coefficients. At the same time, the optics is arranged to preserve chromatic properties, particularly first- and higher-order chromaticities, as much as possible, thereby avoiding additional nonlinear burden on the correction system. The resulting linear optics functions are presented in Figs.~\ref{fig:techoptics}.

\subsection{Straight Sections}
\subsubsection{Damping Wiggler Section}

\begin{figure}[h]
\centering
    \includegraphics[width=1\linewidth]{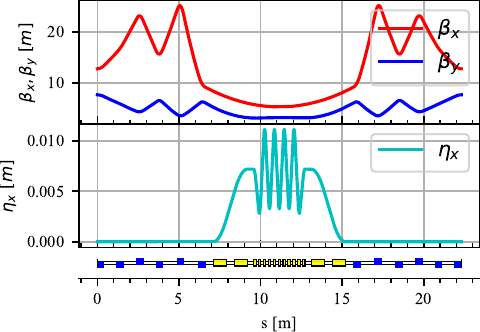}
\caption{Linear optics and lattice layout of the damping-wiggler section for the one-fold STCF.}
    \label{fig:wiggleroptics}
\end{figure}

To provide flexibility in controlling radiation damping time, emittance, and beam parameters at different operating energies, as shown in Fig.~\ref{fig:layout}, two damping wigglers are 
installed in each of the inner and outer half-rings. The lattice and linear optics of the damping-wiggler sections are shown in Fig.~\ref{fig:wiggleroptics}. The wigglers are placed at the center of dedicated straight sections. A controlled non-zero dispersion is generated at the wiggler location by bending magnets positioned on both sides. 

By adjusting the magnetic field strength of these bending magnets, 
the local dispersion at the wiggler can be tuned, allowing controlled 
modification of the horizontal emittance and damping time. This feature is particularly important when the beam energy changes, as it enables compensation of energy-dependent variations in damping time and equilibrium beam size while preserving the overall lattice structure. Six quadrupole families are used for optics matching, ensuring that the working point and beta functions remain essentially unchanged when the wiggler field is varied. In this way, damping optimization can be achieved without perturbing the global tune or chromatic correction scheme.

\subsubsection{Siberian Snake Section}

\begin{figure}[h]
\centering
    \includegraphics[width=1\linewidth]{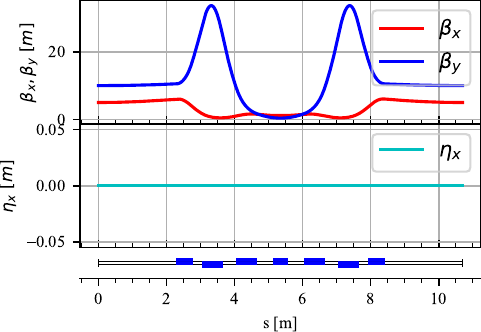}
\caption{Linear optics and lattice layout of the Siberian-snake section in the one-fold STCF.}
    \label{fig:snakeoptics}
\end{figure}

Siberian Snake sections are introduced to achieve longitudinal polarization of the electron beam at the IP. The snakes are distributed around the ring, with equal azimuthal spacing of 120\textdegree\ between neighboring sections. Figure~\ref{fig:snakeoptics} shows the lattice configuration together with the corresponding linear optics functions of a representative snake section.

In the second phase of STCF operation, collisions with polarized electron beam will be implemented, and solenoids will be installed at both ends of each snake section. To preserve longitudinal polarization at the IP and to compensate the transverse coupling and betatron perturbations induced by the solenoidal fields, the optics is carefully matched to realize an $\mathcal{I}$ transformation in the horizontal plane and a $\mathcal{-I}$ transformation in the vertical plane.

\section{\label{sec:globaloptimization}Global nonlinear optimization and lattice performance}

After the IR and technical sections have been individually designed and locally optimized, they are integrated into the full ring, as illustrated in Fig.~\ref{fig:layout}, with the corresponding linear optics shown in Fig.~\ref{fig:ring_optics}. At this stage, the dominant limitations to performance arise from global nonlinear effects, which require coordinated optimization across the entire lattice. The objective is to maximize the dynamic aperture and Touschek lifetime. In this section, we present the global nonlinear optimization strategy for STCF and the resulting performance of the one-fold lattice.

The nonlinear multi-objective optimization uses sextupole and octupole strengths as control variables. The sextupole variables consist of 22 individual or paired magnets, including the IR chromatic sextupoles, mirror-symmetric sextupoles, arc $-\mathcal{I}$ pairs, and HOA sextupoles. Selected IR octupoles are also varied when ADT control is included. Their locations and local functions have been described in Sec.~\ref{sec:ir}.

During the nonlinear optimization, the first-order chromaticities are constrained to small positive values, typically within $1 \lesssim \xi_{x1}, \xi_{y1} \lesssim 3$, to maintain beam stability. With this constraint imposed, a two-stage optimization strategy is then applied, as described in the following subsections.

\subsection{Stage 1: Analysis-driven nonlinear optimization}

The first stage targets the global nonlinear properties that primarily limit the dynamic aperture and Touschek lifetime, while avoiding the computational cost of long-term 6D tracking. Fast surrogate indicators are used, including resonance driving terms evaluated following Ref.~\cite{bengtsson1997sextupole}, the first-order Montague $W$ function, and the chromatic $\beta$-beating coefficients $B_{x1}$ and $B_{y1}$ at selected locations such as the IP and the centers of the crab-waist sextupoles. These quantities provide efficient measures of the dominant nonlinear behavior and allow a broad exploration of the high-dimensional parameter space.

A key control knob in this stage is the phase advance between the IR and arc sections, as well as between adjacent arc modules. These phase relations determine how sextupole-driven nonlinear terms and chromatic beta beating accumulate or cancel around the ring. Using the \texttt{PAMKIT} package~\cite{liu2025PAMKITParticleAcceleratorModeling}, the phase advances in the straight sections are varied using built-in \texttt{Tuning} elements, which introduce pure betatron phase rotations without changing other lattice functions. This enables an efficient scan of phase configurations without repeatedly re-matching the full lattice.

The sextupole strengths are optimized together with these phase advances using a multi-objective genetic algorithm (NSGA-II)~\cite{deb2002FastElitistMultiobjective}. The objectives include minimizing the Montague $W$ function at the IP, reducing chromatic beta beating at the crab-waist sextupoles, maximizing momentum acceptance, and suppressing dominant nonlinear driving terms. The IR octupoles are initially tuned to compensate the leading ADTs, and during the sextupole optimization their strengths are adjusted to keep the detuning coefficients within the target ranges $0 \leq \alpha_{xx} < 1000\ \text{m}^{-1}$, $0 \leq \alpha_{xy} < 2000\ \text{m}^{-1}$, and $0 \leq \alpha_{yy} < 10000\ \text{m}^{-1}$. Once an optimal phase configuration is identified, the virtual \texttt{Tuning} elements are replaced by a physically realizable solution by adjusting quadrupole strengths in the straight sections.

\begin{figure}[h]
    \centering
    \includegraphics[width=1\linewidth]{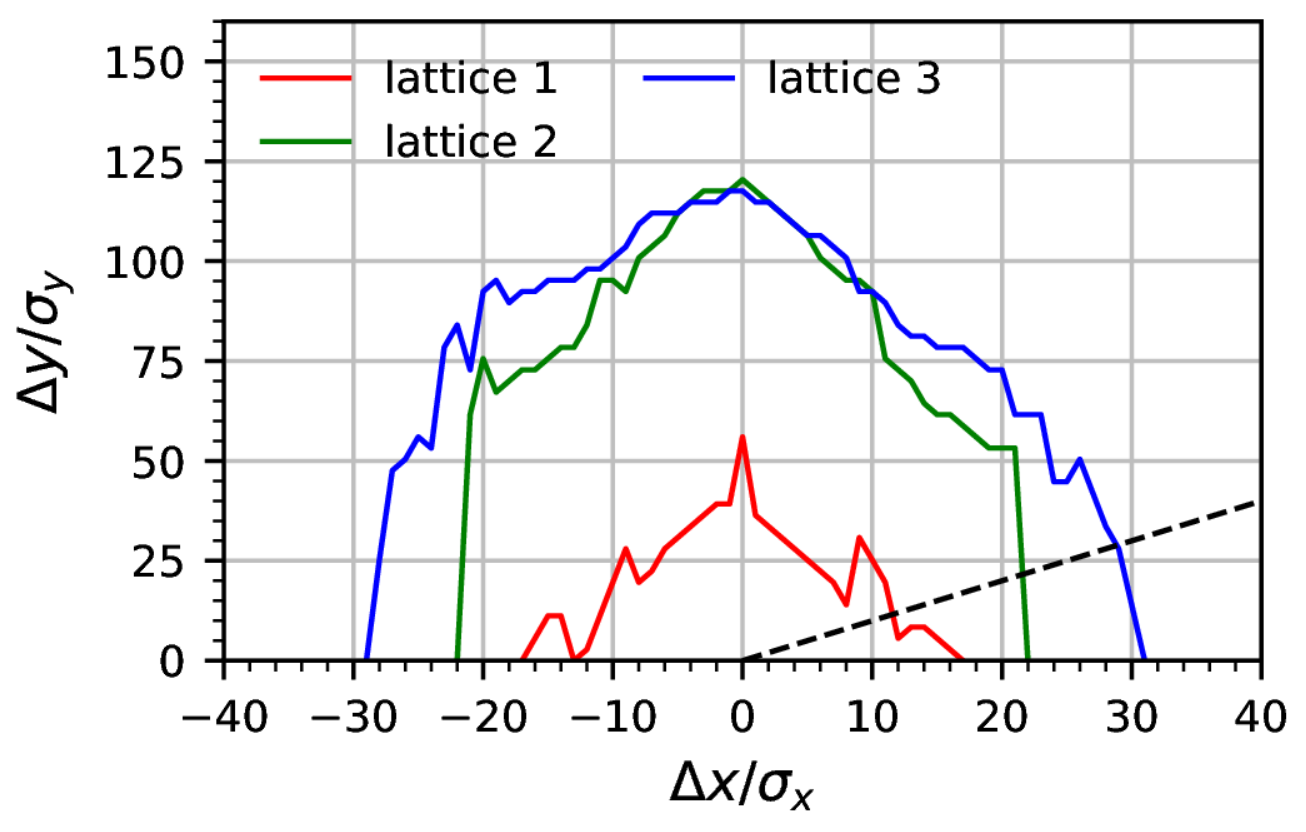}
    \caption{On-momentum dynamic aperture after Stage~1 optimization for the three representative lattices. The dashed black line is defined by $\Delta y/\sigma_y=\Delta x/\sigma_x$.}
    \label{fig:on_DA_after_step1}
\end{figure}
\begin{figure}[h]
    \centering
    \includegraphics[width=1\linewidth]{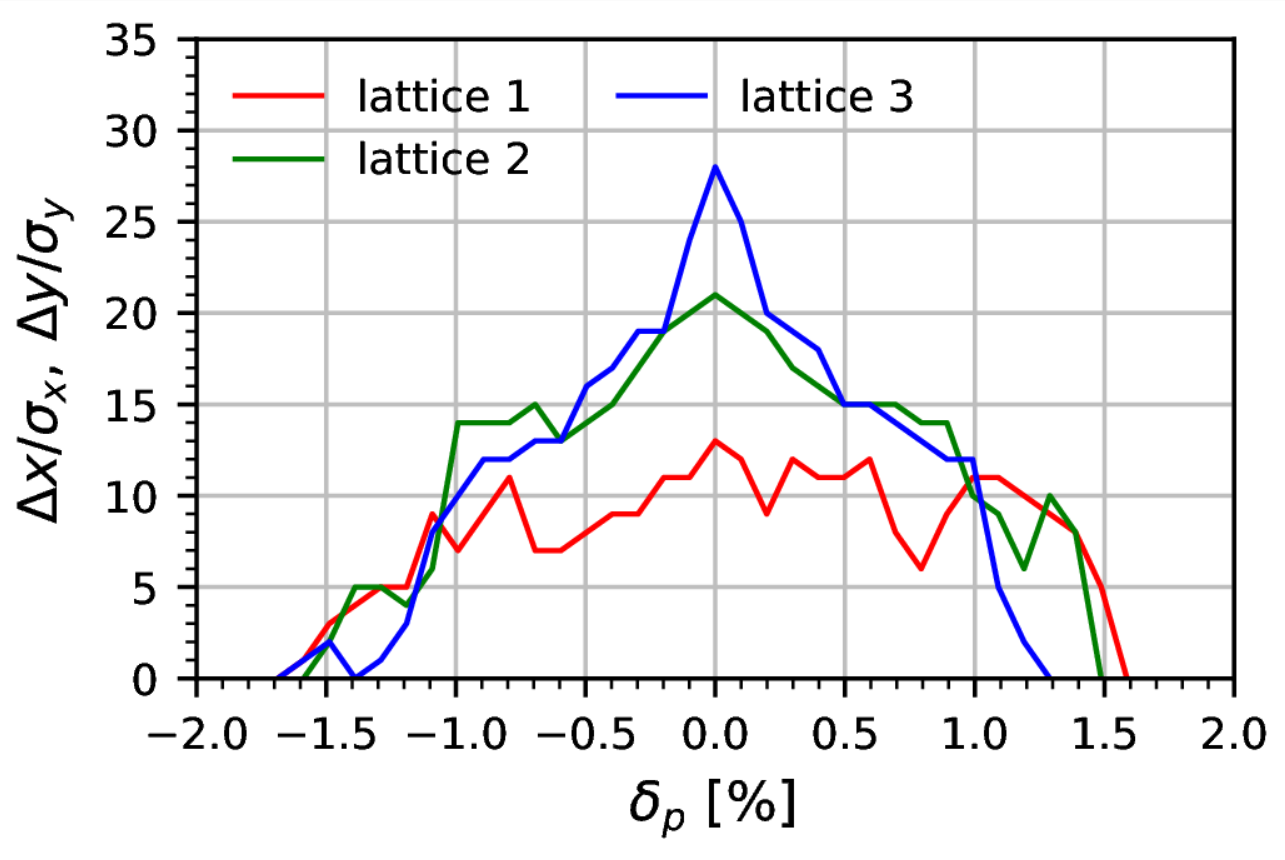}
    \caption{Off-momentum dynamic aperture after Stage~1 optimization for the three representative lattices. The initial transverse coordinates are chosen along $\Delta y/\sigma_y=\Delta x/\sigma_x$, corresponding to the dashed black line in Fig.~\ref{fig:on_DA_after_step1}.}
    \label{fig:off_DA_after_step1}
\end{figure}
\begin{figure}[h]
    \centering
    \includegraphics[width=1\linewidth]{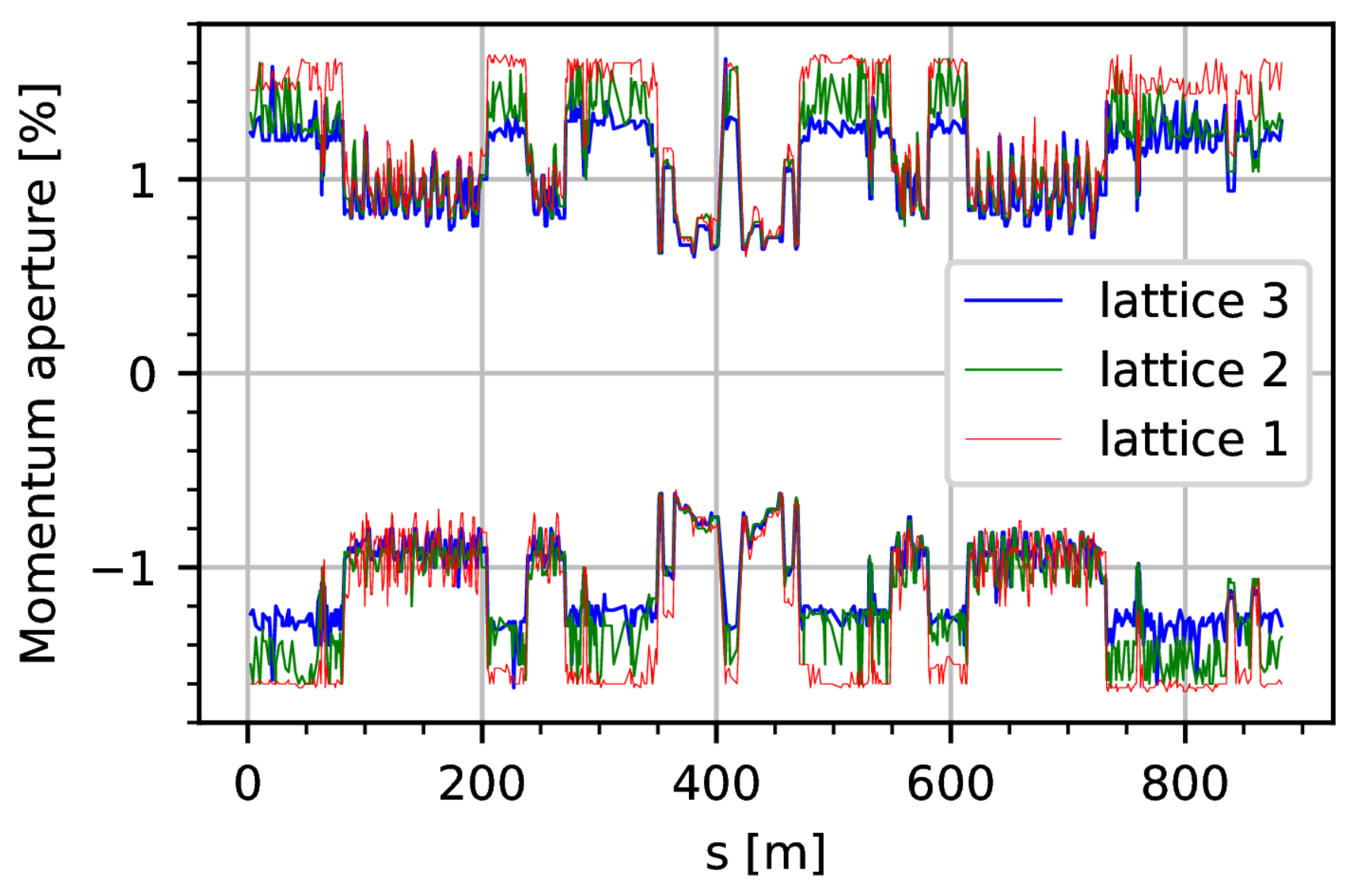}
    \caption{Local momentum acceptance evaluated with \texttt{SAD} for the three representative lattices after Stage~1 optimization.}
    \label{fig:LMA_step1}
\end{figure}

Figures~\ref{fig:on_DA_after_step1}-\ref{fig:LMA_step1} show the on- and off-momentum DA and LMA for three representative lattices obtained in Stage~1: lattice 1 without octupoles, lattice 2 with octupoles activated inside the CW sextupole pair, and lattice 3 with all IR octupoles enabled. As expected, ADT correction by octupoles enlarges the on-momentum DA. However, the additional octupole fields also enhance higher-order chromatic effects, reducing the off-momentum DA and LMA. The corresponding Touschek lifetimes evaluated with \texttt{SAD}~\cite{SADHomePage} are about 180, 320, and 310~s, respectively.

These results expose the main limitation of the analysis-driven stage: surrogate indicators are useful for identifying favorable nonlinear configurations, but they are insufficient to simultaneously optimize the on-momentum DA, off-momentum DA, and LMA. This motivates the tracking-based refinement described below.

\subsection{Stage 2: Tracking-based optimization}

Starting from the Stage~1 solution, Stage~2 directly optimizes lattice performance using full 6D particle tracking. In contrast to Stage~1, which relies on analytical surrogate indicators, this stage evaluates the on-momentum DA, off-momentum acceptance, and Touschek lifetime from tracked particle stability. The phase advances obtained in Stage~1 are kept fixed, while sextupole strengths and, when activated, selected octupole strengths are varied. Unlike in Stage~1, the ADT coefficients are not explicitly constrained; their net impact is assessed through the tracked DA and lifetime. The first-order chromaticities are maintained within the prescribed positive range. Due to the computational cost of tracking, this stage is executed on high-performance computing resources.

To evaluate the Touschek lifetime efficiently, we use the fast Touschek lifetime tracking method~\cite{riemann2024EfficientAlgorithmsDynamic}, in which off-momentum particles are mapped to a reference reduced phase space $(x,p_x,\delta)$ and tested against the tracked stable region. In practice, the simplified implementation available in \texttt{SAD}~\cite{SADHomePage} is used. The tracking model includes exact drift kinematics, Maxwellian fringe fields of the final doublet, and finite-length sextupoles.

The optimization is carried out by varying the strengths of 22 sextupoles (14 in the arcs and 8 in the IR, as described previously), along with selected octupoles when activated, using \texttt{SAD}. To reduce unnecessary tracking evaluations, several preselection criteria are imposed: (i) momentum acceptance exceeding $\pm 1.4\%$, (ii) Montague $W$ function at the IP below 20, (iii) differences of chromatic beta-beatings $B_{x1}$ and $B_{y1}$ between the two crab sextupoles below 20, and (iv) Touschek lifetime from short (100-turn) tracking exceeding 300~s. Only lattices satisfying these conditions are further evaluated with more detailed tracking.

\begin{figure}[h]
    \centering
    \includegraphics[width=1\linewidth]{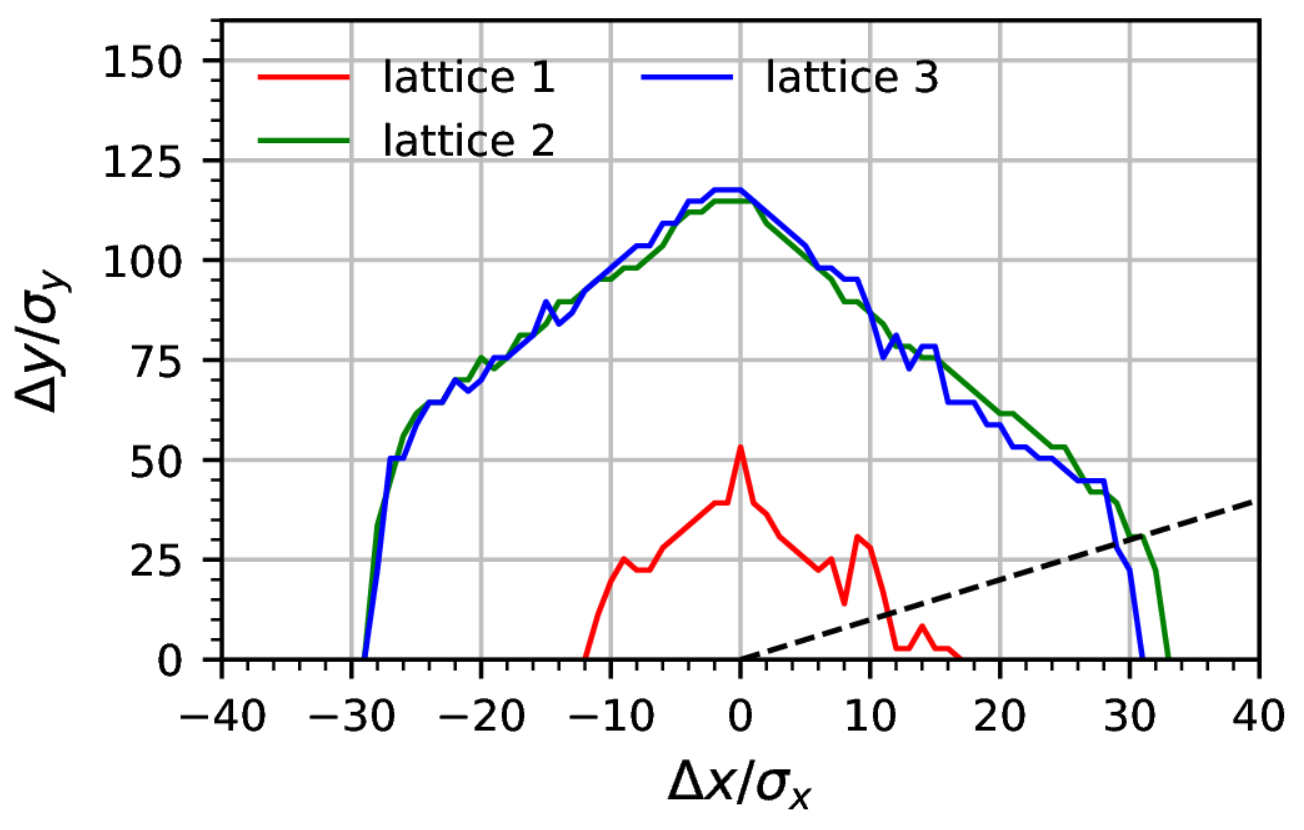}
    \caption{On-momentum dynamic aperture after Stage~2 optimization for the three representative lattices. The dashed black line is defined by $\Delta y/\sigma_y=\Delta x/\sigma_x$.}
    \label{fig:on_DA_after_step2}
\end{figure}
\begin{figure}[h]
    \centering
    \includegraphics[width=1\linewidth]{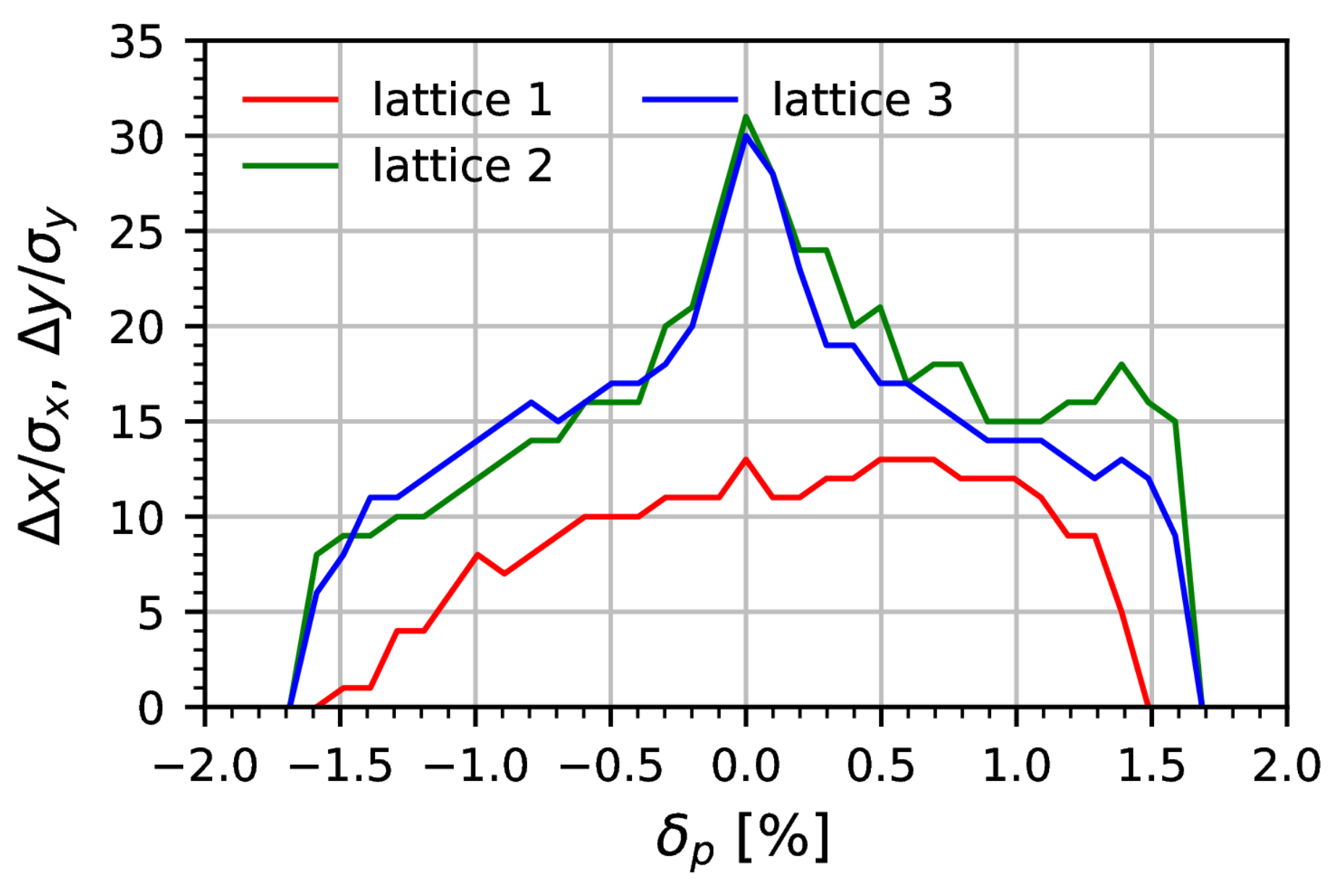}
    \caption{Off-momentum dynamic aperture after Stage~2 optimization for the three representative lattices. The initial transverse coordinates are chosen along $\Delta y/\sigma_y=\Delta x/\sigma_x$, corresponding to the dashed black line in Fig.~\ref{fig:on_DA_after_step2}.}
    \label{fig:off_DA_after_step2}
\end{figure}
\begin{figure}[h]
    \centering
    \includegraphics[width=1\linewidth]{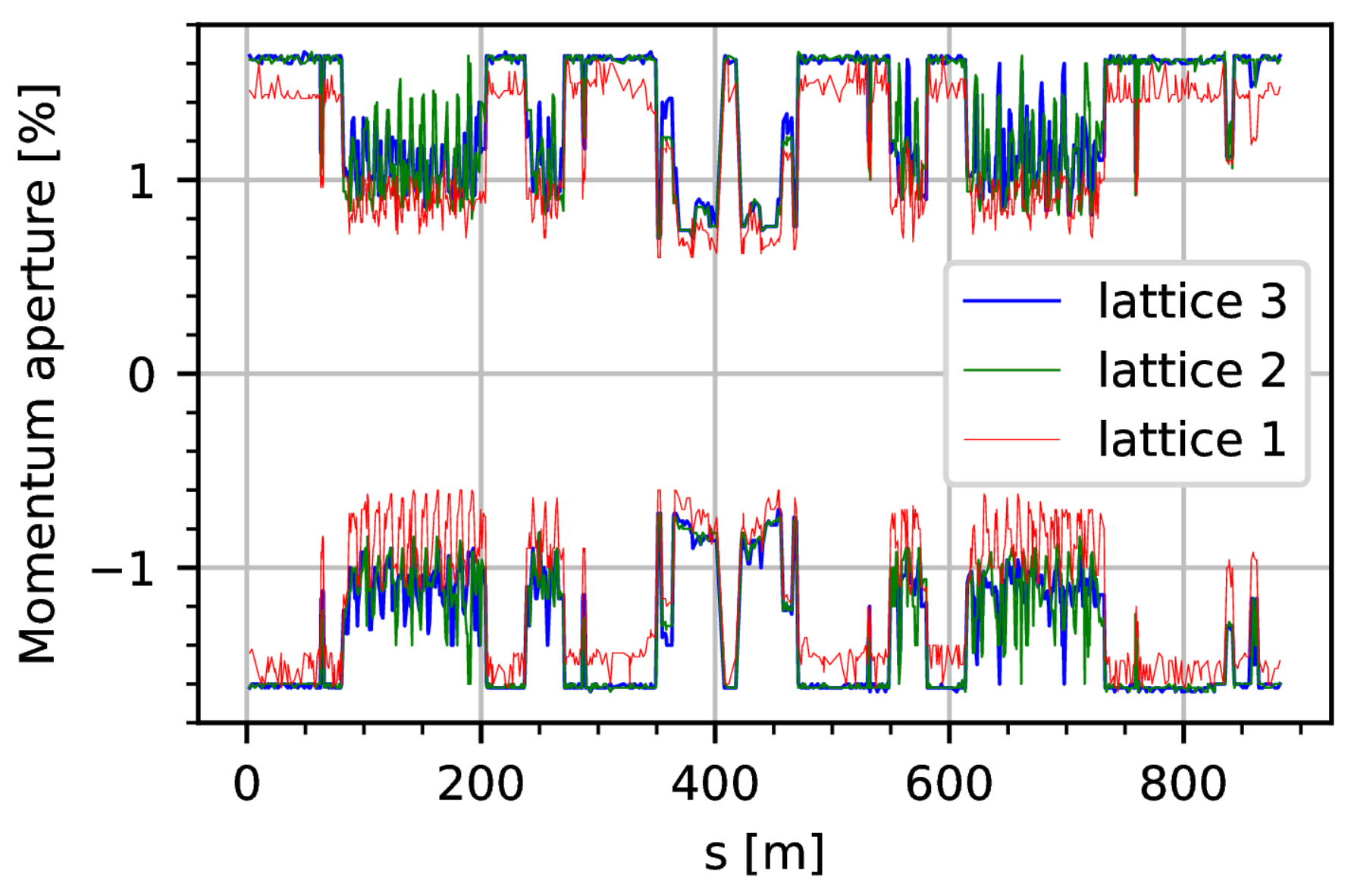}
    \caption{Local momentum acceptance evaluated with \texttt{SAD} for the three representative lattices after Stage~2 optimization.}
    \label{fig:LMA_step2}
\end{figure}

The optimization objectives are to maximize both the on-momentum DA and the Touschek lifetime. The DA is evaluated from 1000-turn tracking, and the lifetime is computed consistently based on the resulting off-momentum aperture. Initial particle coordinates are sampled in transverse phase space along the coordinate $(x,y)$ and momentum $(p_x,p_y)$ directions as
\begin{equation}
X_{1,ij} = \{ n_i \sigma_x, 0, n_i \sigma_y, 0, 0, n_j \sigma_e \},
\end{equation}
\begin{equation}
X_{2,ij} = \{ 0, n_i \sigma_{p_x}, 0, n_i \sigma_{p_y}, 0, n_j \sigma_e \},
\end{equation}
where $n_i$ and $n_j$ denote the transverse and energy scan indices, respectively. The vertical amplitude is scaled with the horizontal amplitude to avoid physically undesirable small vertical apertures during the optimization. The off-momentum DA is defined as the intersection of the stable regions obtained from the coordinate- and momentum-dominated initial conditions.

Figures~\ref{fig:on_DA_after_step2}-\ref{fig:LMA_step2} show the resulting on- and off-momentum DA and LMA for the three representative lattices obtained in Stage~2. Compared with Figs.~\ref{fig:off_DA_after_step1} and~\ref{fig:LMA_step1}, the tracking-based optimization substantially improves both the off-momentum DA and the LMA. The corresponding Touschek lifetimes increase to approximately 260, 540, and 600~s, respectively. These results demonstrate that the adverse off-momentum effects introduced by octupoles can be mitigated when the optimization directly includes tracking-based aperture and lifetime objectives.

\begin{figure*}[htbp]
    \begin{subfigure}[b]{0.48\textwidth}
        \centering
        \includegraphics[width=1\linewidth]{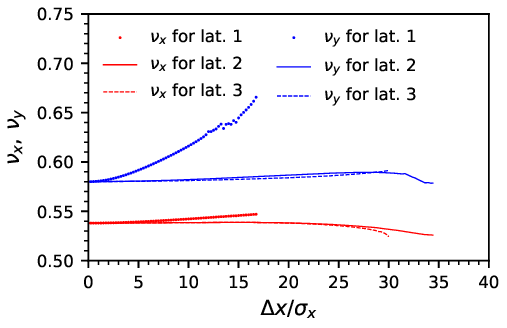}
        \label{fig:tuneshift_wrt_x}
    \end{subfigure}
    \hfill   
    \begin{subfigure}[b]{0.48\textwidth}
        \centering
        \includegraphics[width=1\linewidth]{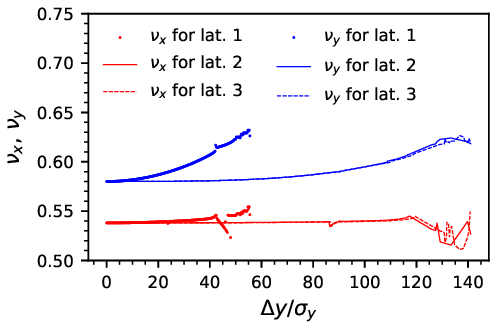}
        \label{fig:tuneshift_wrt_y}
    \end{subfigure}
    \caption{Amplitude-dependent tune shifts with respect to horizontal (left) and vertical (right) amplitudes. Lattice 1, 2, and 3 denote the lattices without octupoles, with octupoles inside the CW sextupole pair, and with octupoles both inside and outside the CW sextupole pair, respectively.}
    \label{fig:all_amplitude_tuneshift} 
\end{figure*}

\begin{figure*}[htbp]
    \centering
    \begin{subfigure}[b]{0.32\textwidth} 
        \raggedleft
        \includegraphics[width=1\linewidth]{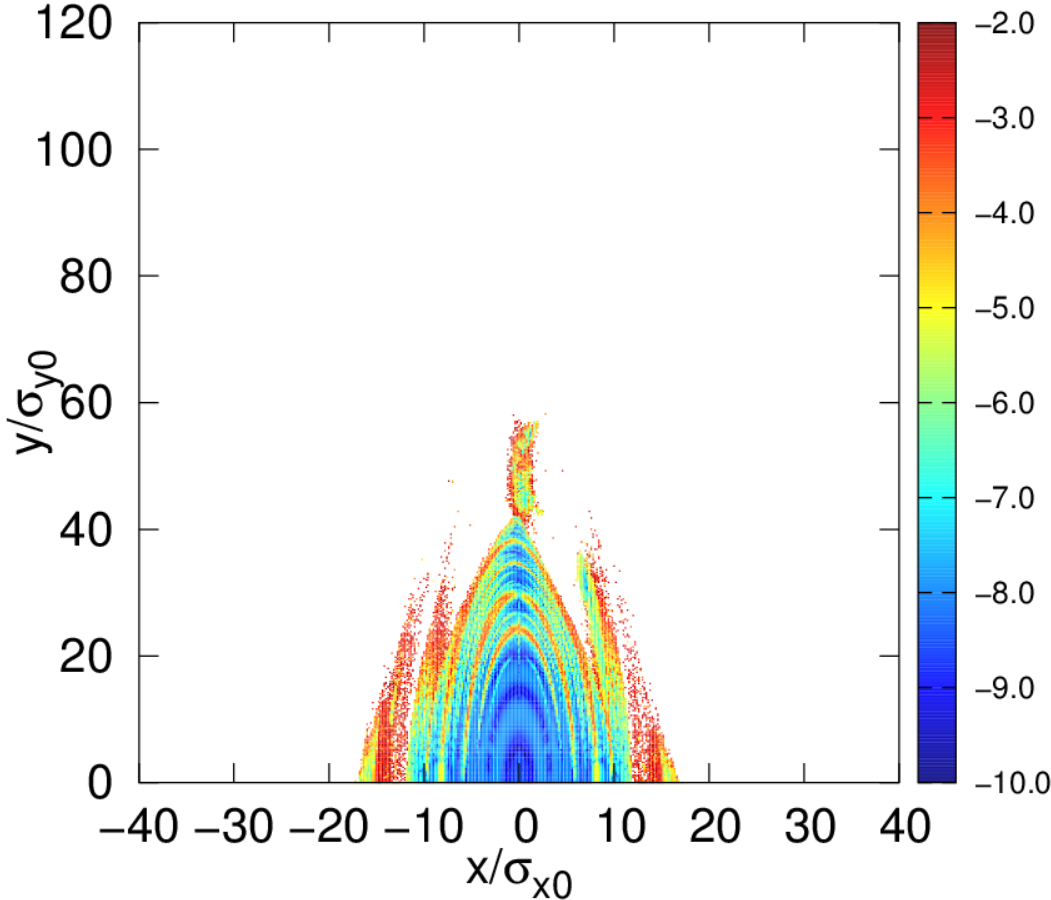}
        \label{fig:XY_FMA_before}
    \end{subfigure}
    \hfill    
    \begin{subfigure}[b]{0.32\textwidth}
        \centering
        \includegraphics[width=1\linewidth]{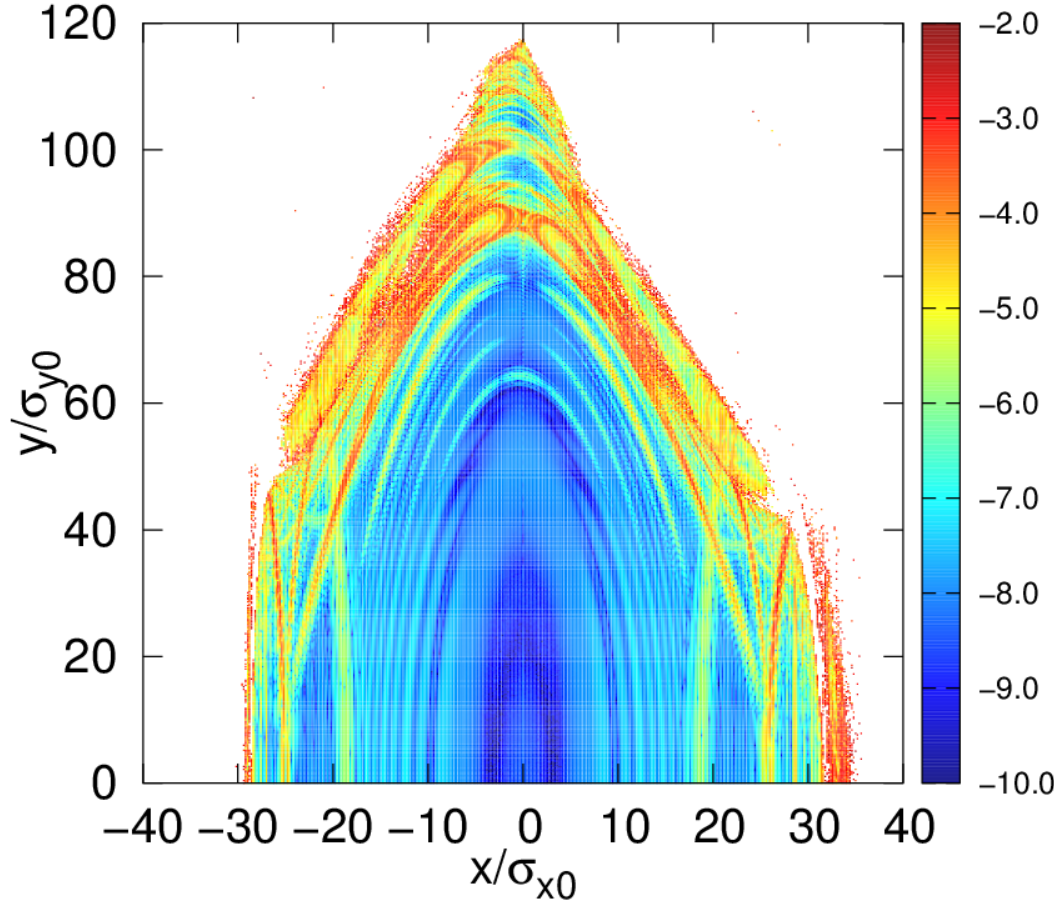}
        \label{fig:XY_FMA_2}
    \end{subfigure}
    \hfill 
    \begin{subfigure}[b]{0.32\textwidth} 
        \raggedleft
        \includegraphics[width=1\linewidth]{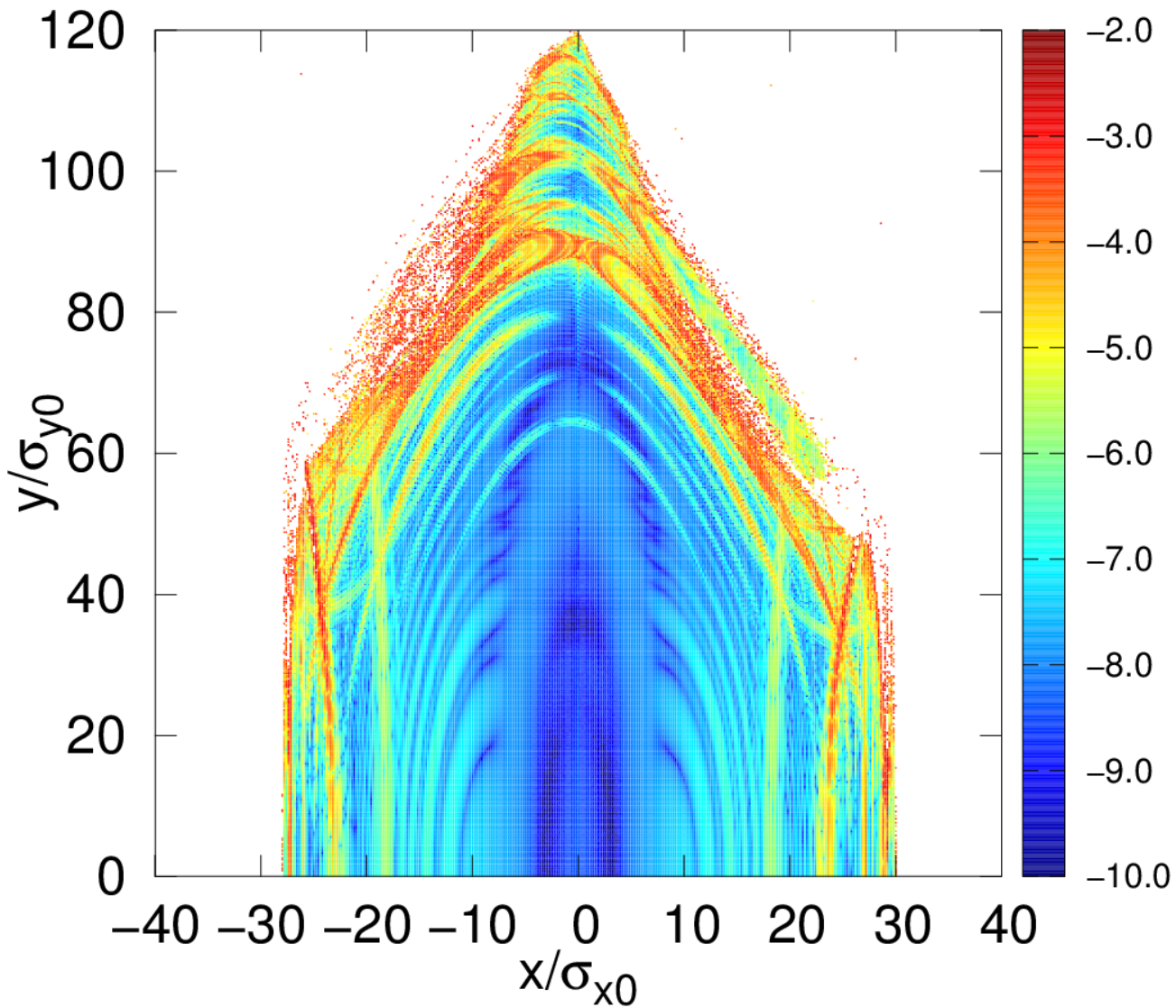}
        \label{fig:XY_FMA_after}
    \end{subfigure}
    \\
    \begin{subfigure}[b]{0.32\textwidth}
        \centering
        \includegraphics[width=1\linewidth]{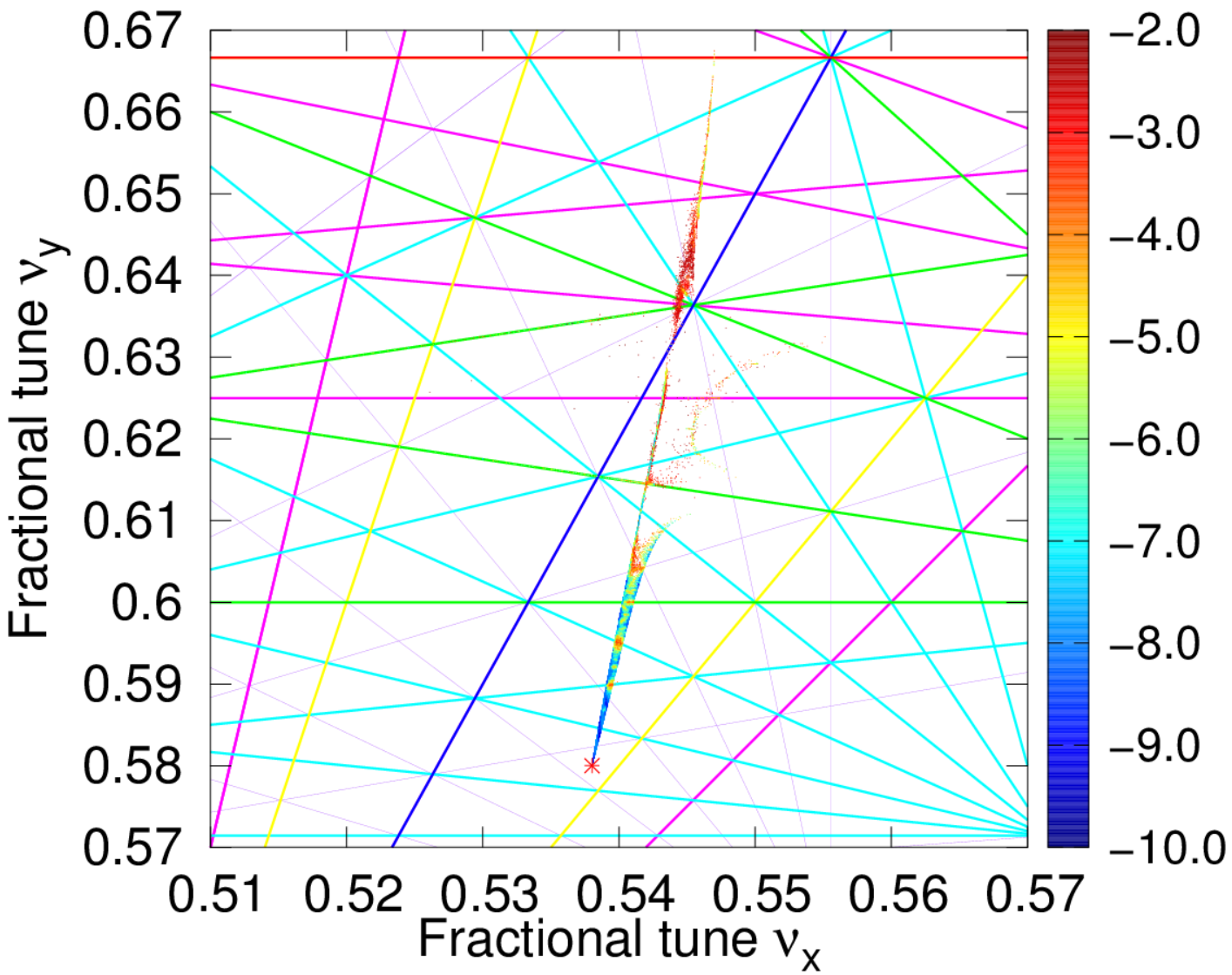}
        \caption{Lattice 1}
        \label{fig:TUNE_FMA_before}
    \end{subfigure}
    \hfill 
    \begin{subfigure}[b]{0.32\textwidth}
        \centering
        \includegraphics[width=1\linewidth]{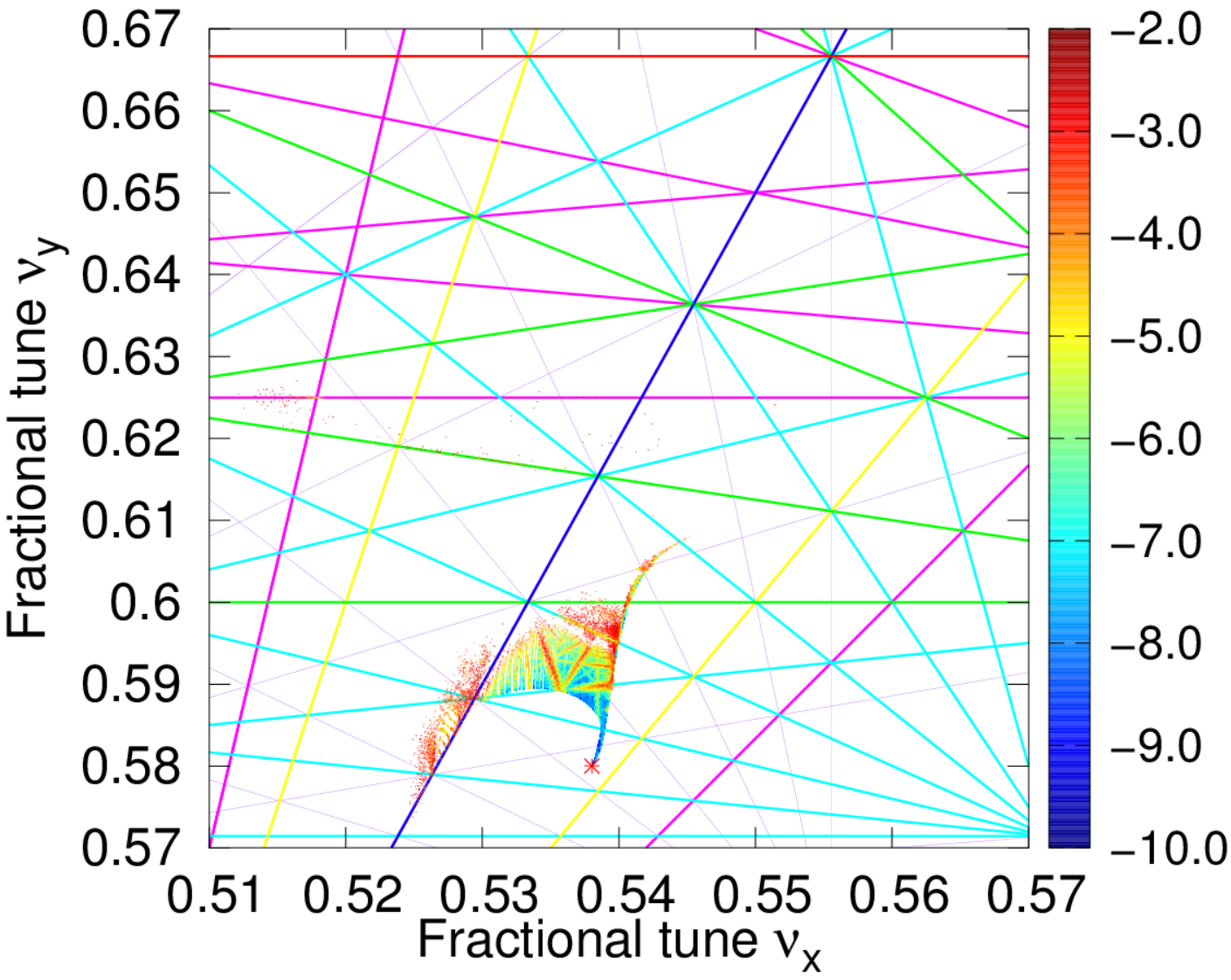}
        \caption{Lattice 2}
        \label{fig:TUNE_FMA_2}
    \end{subfigure}
    \hfill 
    \begin{subfigure}[b]{0.32\textwidth}
        \centering
        \includegraphics[width=1\linewidth]{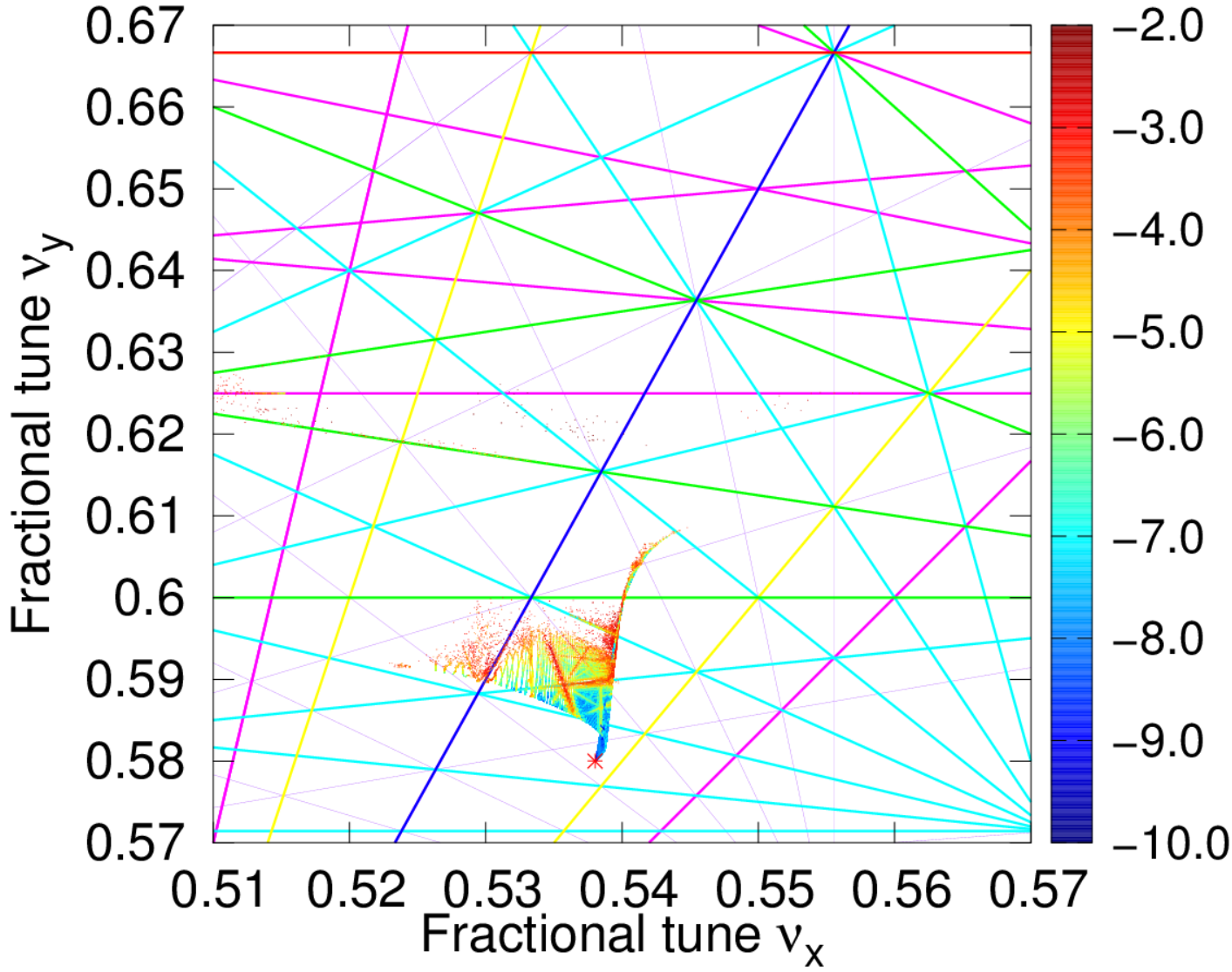}
        \caption{Lattice 3}
        \label{fig:TUNE_FMA_after}
    \end{subfigure}
    \caption{Frequency map analysis for the three Stage~2 lattices. The upper row shows the diffusion rates in transverse coordinate space, and the lower row shows the corresponding tune footprints. The color scale indicates the diffusion rate. From left to right: lattice 1 without octupoles, lattice 2 with octupoles inside the CW sextupole pair, and lattice 3 with all IR octupoles enabled.}
    \label{fig:all_FMA} 
\end{figure*}

To diagnose the role of octupoles in the Stage~2 solutions, Fig.~\ref{fig:all_amplitude_tuneshift} compares the one-dimensional amplitude-dependent tune shifts, while Fig.~\ref{fig:all_FMA} shows the corresponding FMA diffusion maps in the transverse physical space and tune footprints in the betatron-tune space. The octupole-corrected lattices exhibit reduced tune variation with amplitude and a more compact tune footprint, confirming that the improved DA is associated with effective suppression of the leading ADTs.

The comparison also shows that the octupoles outside the CW sextupole pair provide only marginal additional benefit for the present optics. This suggests that, once the dominant detuning inside the CW region is compensated, the need for extra outer octupoles may be reduced. Further studies are nevertheless required to test the robustness of this conclusion.

\section{\label{sec:conclusion}Conclusion and outlook}
In this work, we presented an alternative one-fold lattice design for the STCF collider rings, developed through a unified procedure combining global parameter optimization, LCC-based IR design, and full-ring nonlinear optimization. The optimized design achieves the target luminosity of $10^{35}~\mathrm{cm^{-2}s^{-1}}$ at 2~GeV while maintaining sufficient dynamic aperture, momentum acceptance, and a Touschek lifetime of about 600~s.

A central result is that the nonlinear performance is governed mainly by the transparency of the crab-waist region. The two-stage optimization strategy, consisting of analysis-driven pre-optimization followed by tracking-based refinement, provides an effective way to balance on-momentum DA and off-momentum acceptance. The results also indicate that octupoles inside the CW region are the most effective detuning knobs, whereas additional octupoles outside this region provide only limited further benefit for the present optics.

Future work will address robustness against machine imperfections, the interplay with beam-beam effects, and extension of the method to the nearly two-fold STCF baseline.

\section{Acknowledgments}
T. Liu thanks P. Raimondi for insightful discussions and detailed guidance on crab-waist optics design. This work was supported by the National Key R\&D Program of China under Contract No. 2022YFA1602201, the National Natural Science Foundation of China No.12341501 and No.12405174, and the International Partnership Program of the Chinese Academy of Sciences Grant No. 211134KYSB20200057. We would like to  thank the Hefei Comprehensive National Science Center for their strong support on the STCF key technology research project.

\bibliography{main.bbl}

\end{document}